\documentclass[a4paper,10pt]{article}

\usepackage{cite, amsfonts, amsthm,  fullpage}

\usepackage{amssymb}
\usepackage{amsbsy}
\usepackage{epsfig}
\usepackage{verbatim}

\newcommand{\Pf}{\mathop\mathrm{Pf}\nolimits}
\newcommand{\sgn}{\mathop\mathrm{sgn}\nolimits}
\newcommand{\Pa}{\mathop\mathrm{P}\nolimits}

\newcommand{\bpow}{\mathbf{p}}
\newcommand{\bbpow}{{{\mathbf{p}}^*}}

\theoremstyle{plain}

\newtheorem{Lemma}{Lemma}
\newtheorem{Proposition}{Proposition}

\theoremstyle{remark}
\newtheorem{Remark}{Remark}

\def\l{\langle}
\def\r{\rangle}
\def\g{\Gamma}

\def\Tr{\mathrm {Tr}}
\def\tr{\mathrm {tr}}
\def\det{\mathrm {det}}

\def\res{\mathop{\mathrm {res}}\limits_}

\def\bp{\begin{Proposition}\rm}
\def\ep{\end{Proposition}}
\def\bc{\begin{corollary}}
\def\ec{\end{corollary}}
\def\bl{\begin{Lemma}\em}
\def\el{\end{Lemma}}
\def\be{\begin{equation}}
\def\ee{\end{equation}}
\def\br{\begin{Remark}\rm\small}
\def\er{\end{Remark}}
\def\brs{\begin{remarks}.\\ \rm\
\begin{enumerate}}
\def\ers{\end{enumerate}\end{remarks}}
\def\bea{\begin{eqnarray}}
\def\eea{\end{eqnarray}}

\def\Tr{\mathrm {Tr}}
\def\tr{\mathrm {tr}}
\def\det{\mathrm {det}}

\def\sgn{\mathrm {sgn}}

\def\res{\mathop{\mathrm {res}}\limits}

\def\&{&{\hskip -20pt}}

\def\tp{{\tilde\partial}}
\def\ta{{\tilde\alpha}}

\newcount\YDcount\YDcount=0
\def\YDsize{10pt}

\def\YD#1{%
\ifnum#1=0
 \ifnum\YDcount=0 \ifx\varnothing\undefined\emptyset\else\varnothing\fi
 \else\vskip1.4pt\egroup\YDcount=0\fi
\else
 \ifnum\YDcount=0 \YDcount=1\vcenter\bgroup\vskip1pt
 \else\nointerlineskip\fi
 \vbox{\hrule\hbox{\vrule height\YDsize
 \loop\hskip\YDsize\vrule\ifnum\YDcount<#1\advance\YDcount1\repeat}\hrule
 \kern-0.4pt}\expandafter\YD
\fi}

\begin{document}
\author{ J.W. van de Leur\thanks{Mathematical Institute, Utrecht University, P.O. Box 80010, 3508 TA Utrecht, 
The Netherlands, email: J.W.vandeLeur@uu.nl}\ \and A. Yu.
Orlov\thanks{Institute of Oceanology RAS, Nahimovskii Prospekt 36,
Moscow, Russia,, and National Research University Higher School of Economics, 
International Laboratory of Representation 
Theory and Mathematical Physics,
20 Myasnitskaya Ulitsa, Moscow 101000, Russia, email: orlovs@ocean.ru }}
\title{Character expansion of matrix integrals}

\maketitle

\begin{abstract}

We consider expansions of certain multiple integrals and BKP tau functions in characters of orhtogonal and symplectic groups.
In particular we consider character expansions of integrals over orthogonal and over symplectic matrices.

\end{abstract}

\bigskip 


\section{Introduction}

The character expansion of matrix models used in physics was first presented in the works \cite{DiFIz}, \cite{Kazakov} and was
used in \cite{KMMM}, \cite{Orl}, \cite{HO-2003-Borel}, \cite{HO-2003}, \cite{O-Shiota-2005}, 
\cite{HO-2006} for various problems, in particular in the context of relationships between matrix models and integrable systems.
In all the works mentioned, expansions in terms of Schur functions was used. The importance of such representation
was shown in a set of papers, for instance, for the study of matrix models \cite{Kazakov}, \cite{MirMorSem},\cite{ZinnJustinZuber},
for random processes and random partitions \cite{Adler-van-Moerbeke},\cite{HO-random},
in communications \cite{TulinoVerdu}, \cite{Alfano}
for counting problems, 
see \cite{Okounkov-2000}, \cite{GouldenJackson}, \cite{GGPN}, \cite{AMMN-2014}, \cite{Guay-Harnad-2014}, \cite{NO-LMP}, 
\cite{O-MIHN-2017} 
for relations of quantum and classical models \cite{AntonZabrodin}, \cite{ZabrodinT}, \cite{Zabrodin-}, \cite{FodaWheelZup}, 
\cite{BeteaWheeler} and some others. Here we write down expansions in the characters of orthogonal and symplectic groups. 
We hope that this will also be useful.

\paragraph{Some notations}

Let us recall that the characters of the unitary group $\mathbb{U}(n)$ are labeled by partitions
and coincide with the so-called Schur functions \cite{Mac}. 
A partition 
$\lambda=(\lambda_1,\dots,\lambda_n)$ is a set of nonnegative integers $\lambda_i$ which are called
parts of $\lambda$ and which are ordered as $\lambda_i \ge \lambda_{i+1}$. 
The number of non-vanishing parts of $\lambda$ is called the length of the partition $\lambda$, and will be denoted by
 $\ell(\lambda)$. The number $|\lambda|=\sum_i \lambda_i$ is called the weight of $\lambda$. The set of all
 partitions will be denoted by $\mathbb{P}$.

The Schur function corresponding to $\lambda$ is defined as  the following symmetric function in variables
$x=(x_1,\dots,x_n)$ :
 \be\label{Schur-x}
 s_\lambda(x)=\frac{\det \left[x_j^{\lambda_i-i+n}\right]_{i,j}}{\det \left[x_j^{-i+n}\right]_{i,j}}
 \ee
 in case $\ell(\lambda)\le n$ and vanishes otherwise. One can see that $s_\lambda(x)$ is a symmetric homogeneous 
 polynomial of degree $|\lambda|$ in the variables $x_1,\dots,x_n$.
  
 \br\label{notation} In case the set $x$ is the set of eigenvalues of a matrix $X$, we also write $s_\lambda(X)$ instead
 of $s_\lambda(x)$.
 \er

 There is a different definition of the Schur function as quasi-homogeneous non-symmetric polynomial of degree $|\lambda|$ in 
 other variables, 
 $\bpow =(p_1,p_2,\dots)$, where $\deg p_m = m$:
  \be\label{Schur-t}
  s_\lambda(\bpow)= \det \left[ s_{(\lambda_i-i+j)}(\bpow)\right]_{i,j}
  \ee
  and the Schur functions $s_{(i)}$ are defined by $e^{\sum_{m>0}\frac 1m p_m z^m}=\sum_{m\ge 0} s_{(i)}(\bpow) z^i$.
 The Schur functions defined by (\ref{Schur-x}) and by (\ref{Schur-t}) are equal,  $s_\lambda(\bpow)=s_\lambda(x)$, 
 provided the variables $\bpow$ and $x$ are related by 
  \be
\label{t_m}
  p_m=  \sum_i x_i^m
  \ee
 From now on, we will use  in case the argument of $s_\lambda$ is written as a fat letter  the definition (\ref{Schur-t}),
  and we imply the definition (\ref{Schur-x}) otherwise.

 \br\label{convention}
  For functions $f(\bpow)=f(\bpow(A))$, where
   $p_m(A):= \Tr\, A^m\,, m=1,2,\dots$ and $A$ is a given matrix we may equally write either $f(\bpow(A))$ or
   $f(A)$ where the capital letter implies a matrix.
 In particular under this convention we may write $s_\lambda(A)$ and $\tau(A)$ instead of $s_\lambda(\bpow(A))$ and 
 $\tau(\bpow(A))$.
  
 \er

  \paragraph{Integrals over the unitary group.}
 Consider the following integral over the unitary group which depends on two semi-infinite sets of variables 
 $\bpow=(p_1,p_2,\dots )$ and $\bbpow=(p_1^*,p_2^*,\dots ) $, which are free parameters
 \be
I_{\mathbb{U}(n)}(\bpow,\bbpow):= \int_{\mathbb{U}(n)} 
e^{\tr V\left(\bpow ,U\right) + \tr V\left(\bpow^*,U^{-1}\right)}  d_*U=
 \ee
 \be
\frac{1}{(2\pi )^n} 
\int_{0 \le \theta_1 \le  \dots \le \theta_n\le 2\pi} 
\prod_{1\le j<k\le n}\vert e^{i\theta_j}-e^{-i\theta_k} \vert ^2 
 \prod_{j=1}^n e^{\sum_{m>0}\frac 1m \left(p_me^{im\theta_j} +p_m^* e^{-im\theta_j}\right)}d\theta_j
 \ee
 \be\label{V}
 V(\bpow,x):= \sum_{n>0} \frac 1n p_n x^n
 \ee
 
 Here $d_*U$ is the Haar measure of the group $\mathbb{U}(n)$, see (\ref{Haar-unitary}) in Appendix, and
 $e^{i\theta_1},\dots,e^{i\theta_n}$ are the eigenvalues of $U\in  \mathbb{U}(n)$. The exponential factors
 inside the integral may be treated as a perturbation of the Haar measure and parameters $\bpow, \, \bpow^*$
 are called coupling constants.
 
 Using  the Cauchy-Littlewood identity
 \be
\label{CL}
  \tau(\bpow|\bpow^*):=e^{\sum_{m=1}^\infty \frac 1m p_m^*p_m}=\sum_{\lambda\in \mathbb{P}} s_\lambda(\bpow^*)s_\lambda(\bpow)
  \ee
 and the orthogonality of the irreducible characters of the unitary group
  \be\label{orthonormality-ch-U}
  \int s_\lambda(U)s_\mu(U^{-1})d_*U = \delta_{\lambda,\mu}
  \ee
 we obtain that
 \be\label{Morozov}
 I_{\mathbb{U}(n)}(\bpow,\bbpow) = \sum_{\lambda\in\mathbb{P}\atop
 \ell(\lambda)\le n} s_\lambda(\bpow) s_\lambda(\bbpow)
 \ee
which express the integral over unitary matrices as the "perturbation series in coupling constants".

The formula (\ref{Morozov}) first appeared in \cite{MirMorSem} in the context of the study of Brezin-Gross-Witten model.
It was shown there that the integral $ I_{\mathbb{U}(n)}(\bpow,\bbpow)$ may be related to 
the Toda lattice tau function
of \cite{JM} and \cite{UT} under certain restriction. Then, the series in the Schur functions (\ref{Morozov}) may be related 
to the double Schur functions series found in \cite{Takasaki} and \cite{Takebe}.

In this paper we want to express integrals over the symplectic and over the orthogonal  groups, 
$I_{\mathbb{S}p(N)}(\bpow)$ and $I_{\mathbb{O}(N)}(\bpow)$ respectively, as sums of product
of characters of the orthogonal and of symplectic groups, i.e. to obtain the analogues of the relation (\ref{Morozov}) and 
relate these integrals and sums to integrable systems. On the one hand we shall relate $I_{\mathbb{S}p(N)}(\bpow)$ 
and $I_{\mathbb{O}(2n)}(\bpow)$  to the DKP\footnote{We need to note that the DKP hierarchy has other names. 
It was rediscovered in \cite{AvM-Pfaff} using the approach different of \cite{JM} and called Pfaff lattice. 
It was also called coupled KP equation in \cite{HiO}}, and we shall relate $I_{\mathbb{O}(2n+1)}(\bpow)$ to BKP tau functions, 
introduced respectively in \cite{JM} and \cite{KvdLbispec} and obtain Pfaffian representation for these integrals. On the other
hand one can relate these integrals to the Toda lattice (TL) tau function \cite{JM},\cite{UT} which yields the determinant 
representation.

We show that the so-called $\beta=1,2,4$ ensembles may be written as formal series in characters.

\section{Polynomials $o_\lambda(\bpow)$ and $sp_\lambda(\bpow)$ and TL tau functions $\tau_\pm(\bpow|\bpow^*)$}

The orthogonal and symplectic characters are also labeled by partitions. They are given by the following expressions

\be
\label{o}
o_\lambda=
\frac{\det \left[ x_j^{\lambda_i+n-i+1}-x_j^{-\lambda_i-n+i}\right]_{1\le i,j\le n}}
{\det \left[ x_j^{n-i+1}-x_j^{-n+i}\right]_{1\le i,j\le n}}
\ee
and
\be
\label{sp}
sp_\lambda=
 \frac{\det \left[ x_j^{\lambda_i+n-i+1}-x_j^{-\lambda_i-n+i-1}\right]_{1\le i,j\le n}}
 {\det \left[ x_j^{n-i+1}-x_j^{-n+i-1}\right]_{1\le i,j\le n}}
\ee
respectively. See \cite {FH} or Appendix \ref{Characters of} for more information.
Baker \cite{Baker} realized that these characters can be obtained from the corresponding Schur functions $s_\lambda$ 
by action of some operator. For this, it will be convenient to use the Schur functions in terms of the power sums $p_m$.
As usual, write $\tilde\partial=(\partial_{p_1},2 \partial_{p_2}, 3 \partial_{p_3},\ldots)$. Let 
\be
\label{Omega}
\Omega_\mp(\bpow)= \sum_{m>0}\left(-\frac{1}{2m} p_m^2 \mp  \frac{1}{2m} p_{2m}\right)\,,\quad
\Omega_\mp=\Omega_\mp(\tp) :=\sum_{m>0}\left(-\frac {m}{2} (\partial_m)^2 \mp  \partial_{2m}\right)
\ee
then
\be
\label{schur-to o or sp}
o_\lambda(\bpow) = e^{\Omega_-(\tp)} \cdot s_\lambda(\bpow)\,,
\quad 
sp_\lambda(\bpow)= e^{\Omega_+(\tp)} \cdot s_\lambda(\bpow) 
\ee
Hence, if we let the operator $ \Omega_\mp(\tp^*)$ act on the Cauchy- Littelewood identity (\ref{CL}), we obtain
  \be\label{series-in-characters-orth}
  \tau_-(\bpow|\bpow^*)=\sum_\lambda o_\lambda (\bpow^*) s_\lambda(\bpow)
  \ee
  and
  \be\label{series-in-characters-sympl}
  \tau_+(\bpow|\bpow^*)=\sum_\lambda sp_\lambda (\bpow^*) s_\lambda(\bpow)
  \ee
where
\be\label{f1}
 \tau_\mp(\bpow|\bpow^*)=e^{-\frac 12\sum_{m=1}^\infty\,\frac 1m p_m^2\,\mp\,\sum_{m=1}^\infty \, \frac{1}{2m}{ p}_{2m} 
  +\sum_{m=1}^\infty \, \frac 1m{ p}_m p_m^*}
\ee

 \br \label{via-TL-vac}
Note that
\be
	\tau_\mp(\bpow|\bpow^*)=e^{\Omega_\mp(\bpow)} \tau_0(\bpow|\bpow^*)  
\ee
where $ \tau_0(\bpow|\bpow^*) =e^{\sum_{m>1} \frac 1m p_mp_m^*}$ is known to be the simplest tau function of the TL hierarchy 
(this simplest tau function does not depend
on the discrete TL time $p_0$).
\er

 It is well known that the function $ \tau_0(\bpow|\bpow^*)$, for the variables $\bpow=(p_1,p_2,\ldots)$, is a solution 
 of the  Hirota  bilinear equations for the KP hierarchy:
 \be
\label{KP}
  \oint\frac{dz}{2\pi i} e^{V(\bpow'-\bpow,z)}
  \tau_0(\bpow'-[z^{-1}]|\bpow^*)
  \tau_0(\bpow +[z^{-1}]|\bpow^*) =0
 \ee
Here  $V$ is given by (\ref{V}) and the variables  $\bpow^*=(p_1^*,p_2^*,\ldots)$ play the role of auxiliary parameters.
Here and below the notation $[a]$ serves to denote the following set of power sums: $\left(a,a^2,a^3,\ldots  \right)$.
The action of $e^{\Omega_\mp(\tp^*)}  e^{\Omega_\mp(\tilde\partial^{\prime *} )}$ on (\ref{KP}) gives 
 \be
\label{KP2}
  \oint\frac{dz}{2\pi i} e^{V(\bpow'-\bpow,z)}
  \tau_\mp(\bpow'-[z^{-1}]|\bpow^*)
  \tau_\mp(\bpow +[z^{-1}]|\bpow^*) =0
 \ee
hence $ \tau_\mp(\bpow|\bpow^*)$  is also a tau function of the KP hierarchy. Then it follows from Remark \ref{via-TL-vac}
that both $\tau_\pm(\bpow|\bpow^*)$ are TL tau functions where $\bpow$ and $\bpow^*$ are two sets of the higher times. 
These tau functions do not depend on the discrete TL variable $t_0$ because $\tau_0(\bpow|\bpow^*)$ does not depend on it. 

 According to Sato \cite{Sato} a KP tau function may be related to an element of an infinite
  dimensional Grassmannian as a series in the Schur functions 
  \[
   {\rm tau}^{\rm KP}(\bpow) = \sum_\lambda \pi_\lambda s_\lambda(\bpow)
  \]
  where $\pi_\lambda$ are the Pl\"ucker coordinates of the element. Hence according to (\ref{CL}), (\ref{series-in-characters-orth}) 
  and (\ref{series-in-characters-sympl}), the functions  $s_\lambda(\bpow^*)$, $o_\lambda(\bpow^*)$ and $sp_\lambda(\bpow^*)$ are the Pl\"ucker 
  coordinates of $\tau_0(\bpow|\bpow^*)$, $\tau_-(\bpow|\bpow^*)$ and $\tau(\bpow|\bpow^*)$, respectively.
  The related elements of the Grassmannian is written down in Appendix \ref{characters}.
  
The Pl\"ucker coordinates $o_\lambda(\bpow^*)$ and $sp_\lambda(\bpow^*)$ may be evaluated respectively as follows
\be\label{o-explicit}
o_\lambda (\bpow^*) = \left( s_\lambda({\tilde \partial})\cdot 
e^{-\frac 12\sum_{m=1}^\infty\,\frac 1m { p}_m^2\,-\,\sum_{m=1}^\infty \, \frac{1}{2m}{ p}_{2m} 
  +\sum_{m=1}^\infty \, \frac 1m { p}_m p_m^*}
\right)\vert _{\bpow=0}
\ee
and
\be\label{sp-explicit}
sp_\lambda (\bpow^*) = \left( s_\lambda({\tilde \partial})\cdot 
e^{-\frac 12\sum_{m=1}^\infty\,\frac 1m{ p}_m^2\,+\,\sum_{m=1}^\infty \, \frac{1}{2m}{ p}_{2m} 
  +\sum_{m=1}^\infty \, \frac 1m { p}_m p_m^*}
\right)\vert_{\bpow=0}
\ee
which may be compared with the identity for the Schur functions
\be
s_\lambda (\bpow^*) = \left( s_\lambda({\tilde \partial})\cdot 
e^{\sum_{m=1}^\infty \, \frac 1m{ p}_m p_m^*}
\right)\vert_{\bpow=0}
\ee

As in the previous section, let us assign the weight $k$ to $p_k$. We recall that the polynomials $s_\lambda$ are 
quasi-homogeneous in the variables $p_m$ 
of the weight $|\lambda|$.  
As we see from (\ref{o-explicit}) and (\ref{sp-explicit}) polynomials $o_\lambda$ and $sp_\lambda$ are not quasi-homogeneous:
they both may be presented as $s_\lambda$ plus polynomials of minor weights.

For instance
\[
 o_{(1)} (\bpow)=sp_{(1)}(\bpow) = s_{(1)}(\bpow) =p_1
 \]
 \[
 o_{(2)}(\bpow) = s_{(2)}(\bpow)-1=\frac 12 p_2+\frac12 p_1^2-1\,,
 \quad sp_{(2)}(\bpow) = s_{(2)}(\bpow)=\frac 12 p_2+\frac12 p_1^2
\]
\[
 o_{1^2}(\bpow) = s_{1^2}(\bpow)=-\frac 12 p_2+\frac12 p_1^2\,,
 \quad sp_{1^2}(\bpow) = s_{1^2}(\bpow) -1=-\frac 12 p_2+\frac12 p_1^2 -1
\]

Next from
$
 \tau_-(-\bpow|-\bpow^*) =  \tau_+(\bpow|\bpow^*)
$
and from  $s_\lambda(\bpow)=(-)^{|\lambda|}s_{\lambda^{\rm tr}}(-\bpow)$,  we get
\be\label{duality}
 sp_\lambda(\bpow)=(-)^{|\lambda|}o_{\lambda^{\rm tr}}(-\bpow)
\ee
where $-\bpow =(-p_1,-p_2,-p_3,\dots )$.

From the following well-known formulas (see \cite{Mac}, pages 76 and 77 or \cite{Littlewood} page 238)
\[
e^{\frac12 \sum_{m>0} \frac 1m p^2_m+\sum_{m>0,{\rm odd}}\frac 1m p_m}=\sum_{\mu\in\Pa}\, s_{\mu}(\bpow)\, ,\quad 
 e^{-\Omega_+(\bpow)}=\sum_{\mu\in\Pa}\, s_{\mu\cup\mu}(\bpow)\,,\ \mbox{and }
 e^{-\Omega_-(\bpow)}=\sum_{\mu\in\Pa_{\rm even}} s_{\mu}(\bpow)
\]
where $\Pa_{\rm even}$
 is the set of all partitions with even parts (including $(0)$), one deduces
 \bl \label{exponentials}{\em
 \bea\label{SchurSum1''}
\sum_{\mu\in\Pa}\,z^{|\mu|}s_\mu({\tilde\partial})\,&=&\,
e^{\frac 12\sum_{m=1}^\infty\, m z^{2m} \partial_m^2\,+\,\sum_{m>0,{\rm odd}} \, z^m \partial_m}
\\
 \label{SchurSum4''}
\sum_{\mu\in\Pa}\,z^{2|\mu|} s_{\mu\cup\mu}({\tilde\partial})\,&=&\,
e^{\frac 12 \sum_{m=1}^\infty\,m z^{2m} \partial_m^2\,-\,\sum_{m>0,\,{\rm even}} \, z^m \partial_m}
= e^{-\Omega_{+}(z^m\tp_m)}
 \\
 \label{SchurSum-even''}
\sum_{\mu\in\Pa_{\rm even}}\,z^{|\mu|} s_{\mu}({\tilde\partial})\,&=&\,
e^{\frac 12 \sum_{m=1}^\infty\,m z^{2m} \partial_m^2\,+\,\sum_{m>0,\,{\rm even}} \,z^m \partial_m }
= e^{-\Omega_{-}(z^m\tp_m)}
\eea
 where $s_\lambda({\tilde\partial})$ is defined as in (\ref{o-explicit})-(\ref{sp-explicit}).
 }
\el
(see also \cite{OST-I} where in (\ref{SchurSum4''}) and (\ref{SchurSum-even''}) there is the opposite sign for linear term 
in exponents which is a misprint).

\br
From Lemma \ref{exponentials}
a number of relations may be obtained. We present two examples:
 \[
 e^{- \sum_{m>0} \frac{1}{2m} p_m^2 -\sum_{m>0,\,{\rm odd}} \frac 1m p_m}
\sum_{\mu\in\Pa} s_{\mu/\lambda}(\bpow)=
\sum_{\mu\in\Pa} s_{\lambda/\mu}(\bpow)= 
e^{\sum_{m=1}^\infty\,(m \partial_m^2\,+\, \partial_m)  } \cdot o_\lambda(\bpow)
\]
where the first equality is obtained from Ex 27(a) in I.5 of \cite{Mac}. The second example follows from
(\ref{schur-to o or sp}), (\ref{SchurSum1''}) and from $s_\mu({\tilde\partial})\cdot s_\lambda(\bpow)=s_{\lambda /\mu}(\bpow)$.
The second example, 
\be
\label{last}
\sum_{\mu\in\Pa} s_{\lambda/\mu\cup\mu}(\bpow) = e^{\sum_{m>1} m \partial_m^2}\cdot  o_\lambda(\bpow)
 \ee
 we obtain from (\ref{schur-to o or sp}) and (\ref{SchurSum4''}).
Note, that the constant term
$\left[ e^{\sum_{m>1} m \partial_m^2}\cdot  o_\lambda(\bpow) \right]_{\bpow =0}$
 of  (\ref{last}) 
 is equal to $1$ for any  
$\lambda$ of form $\mu\cup\mu$, and vanishes otherwise.
 
\er

\paragraph{Relation to irreducible characters of the orthogonal and symplectic groups.}

We shall use notations explained in Remark \ref{convention} with $p_m(U)=\Tr\, U^m$.

 In this notation we write
 \be\label{f+t*}
 \tau_+(U|\bpow^*)= \prod_{i < j}(1 - x_ix_j)\prod_{k=1}^n e^{\sum_{m=1}^\infty \frac 1m p^*_m x^m_k} =
 \sum_\lambda sp_\lambda (\bpow^*) s_\lambda(U)
 \ee
 \be\label{f-t*}
 \tau_-(U|\bpow^*)= \prod_{i \le j}(1 - x_ix_j)\prod_{k=1}^n e^{\sum_{m=1}^\infty \frac 1m p^*_m x^m_k} =
 \sum_\lambda o_\lambda (\bpow^*) s_\lambda(U)
 \ee
 where $x_1,\dots,x_n$ are the eigenvalues of $U\in\mathbb{U}(n)$.
 Let us note that
 \be\label{f-=f+det(1-U2)}
 \tau_-(U|\bpow^*) = \tau_+(U|\bpow^*)\det(1-U^2) 
 \ee
 Now take  $Z\in\mathbb{S}p(2n)$ and let
 $z_1,z_1^{-1},\dots,z_n,z_n^{-1}$ be  the eigenvalues of $Z$. Then the formula  (\ref{f+t*}) reads
 \be\label{series-in-sp-s'}
 \prod_{i < j\le n}(1 - x_ix_j)\prod_{i,j=1}^n (1 - x_iz_j)^{-1} (1 - x_iz_j^{-1})^{-1}=
\sum_{\lambda\in\mathbb{P}}  sp_\lambda (Z) s_\lambda(U)\,,\quad Z\in\mathbb{S}p(2n),\,\, U\in\mathbb{U}(n)
 \ee
 This relation is known as the Cauchy identity for the irreducible characters of $\mathbb{S}p(2n)$, see \cite{Littlewood}.
 Thus, from the completeness of the Schur functions $s_\lambda$ in the space of the symmetric functions
 in $x_1,\dots, x_n$, it follows that the polynomials $sp_\lambda(\bpow(Z))=sp_\lambda(Z)$ coincide with the 
 characters of $\mathbb{S}p(2n)$.
 
 Similarly, for $Z\in\mathbb{O}(2n)$ with the eigenvalues $z_1,z_1^{-1},\dots,z_n,z_n^{-1}$, and for
 $U\in\mathbb{U}(n)$ with eigenvalues $x_1,\dots, x_n$, we obtain
 \be\label{series-in-o-s'-2n}
 \prod_{i\le j\le n}(1 - x_ix_j)\prod_{i,j=1}^n (1 - x_iz_j)^{-1}(1 - x_iz_j^{-1})^{-1} =
\sum_{\lambda\in\mathbb{P}}  o_\lambda (Z) s_\lambda(U)\,,\quad Z\in\mathbb{O}(2n),\,\, U\in\mathbb{U}(n)
 \ee
 While for $Z\in\mathbb{O}(2n+1)$ with the eigenvalues $z_1,z_1^{-1},\dots,z_n,z_n^{-1},1$, we obtain
  \be\label{series-in-o-s'-2n+1}
 \prod_{i\le j\le n}(1 - x_ix_j)\prod_{i,j=1}^n (1 - x_iz_j)^{-1}(1 - x_iz_j^{-1})^{-1}
 \prod_{i=1}^n (1 - x_i)^{-1}=
\sum_{\lambda\in\mathbb{P} } o_\lambda (Z) s_\lambda(U)\,,\quad Z\in\mathbb{O}(2n+1),\,\, U\in\mathbb{U}(n)
 \ee
 Relations (\ref{series-in-o-s'-2n}), (\ref{series-in-o-s'-2n+1}) are known as
 the Cauchy identities for the orthogonal group, the polynomials $o_\lambda(Z)$
 are irreducible characters of the orthogonal group.
 
 We have
 \[
  \tau_+(U|Z)=  \sum_\lambda sp_\lambda (Z) s_\lambda(U)\,,\quad Z\in\mathbb{S}p(2n),\quad U\in\mathbb{U}(n)
 \]
and
 \[
  \tau_-(U|Z)=\sum_\lambda o_\lambda (Z) s_\lambda(U)\,,\quad Z\in\mathbb{O}(N),
  \quad U\in\mathbb{U}(n),\, n=\left[\frac N2\right]
 \]
 where  $sp_\lambda$ and $o_\lambda$ are characters respectively of symplectic and orthogonal groups.
 
The content of this Section may be compared with \cite{KFG}, \cite{Baker} where universal characters of classical 
groups were considered.

\section{Characters and fermions\label{characters-and-fermions}}

Since all Schur functions $s_\lambda$ are in the $GL_\infty$ group orbit, they are KP tau functions, i.e., they satisfy the 
bilinear identity:
\be 
\label{KP'}
{Res}_z \psi(z)\tau\otimes \psi^\dagger(z)\tau=0
\ee
where $\psi(z)=\sum_{i\in\mathbb{Z}}\psi_i z^i$ and $\psi^\dagger (z)=\sum_{i\in\mathbb{Z}}\psi_i^\dagger z^{-i-1}$,
are  free fermionic fields (see  \cite{JM}), 
 whose Fourier components 
anti-commute as follows
$\psi_i\psi_j+\psi_j\psi_i=\psi^\dag_i\psi^\dag_j+\psi^\dag_j\psi^\dag_i=0$ and 
$\psi_i\psi^\dag_j+\psi^\dag_j\psi_i=\delta_{i,j}$ where $\delta_{i,j}$ is the Kronecker symbol. We put
 \be
\psi_i|0\r=\psi^\dag_{-1-i}|0\r =\l 0|\psi_{-1-i}=\l 0|\psi^\dag_{i}=0, \qquad\mbox{for } i<0,
 \ee
where $\l 0|$ and $|0\r$ are left and right vacuum vectors of the fermionic Fock space, $\l 0|\cdot 1 \cdot |0\r=1$. 
 Let 
\be
\l n|=
\cases{
\l 0|\psi^\dag_{0}\cdots \psi^\dag_{n-1}\quad {\rm if }\quad n > 0 \cr
\l 0|\psi_{-1}\cdots \psi_{-n}\quad {\rm if }\quad n < 0 
 }
\,,\quad
|n \r =
\cases{
\psi_{n-1}\cdots \psi_{0}|0\r\, \quad {\rm if }\quad n > 0 \cr
\psi^\dag_{-n}\cdots \psi^\dag_{-1}|0\r\, \quad {\rm if }\quad n < 0 
 }
 \ee
then $\l n|\cdot 1 \cdot |m\r=\delta_{n,m}$.
Note that $e^{\Omega_\mp}$ is an automorphism of the Fock space. It maps any charge sector into itself and maps Schur functions 
into orthogonal and symplectic characters, see (\ref{schur-to o or sp}).
Thus $o_\lambda$ (for $e^{\Omega_-}$) and $sp_\lambda$ (for $e^{\Omega_+}$) satisfies
\be 
\label{KP2-fermi-Hirota}
{Res}_z e^{\Omega_\mp(\ta)}\psi(z)e^{-\Omega_\mp(\ta)}\sigma\otimes 
e^{\Omega_\mp(\ta)}\psi^\dagger(z)e^{-\Omega_\mp(\ta)}\sigma=0,
\ee
where $\sigma=
e^{\Omega_\mp}\tau$.
We now want to calculate $\Psi_\mp (z)=e^{\Omega_\mp(\ta)}\psi(z)e^{-\Omega_\mp(\ta)}$ and 
$ \Psi_\mp^\dagger (z)=e^{\Omega_\mp(\ta)}\psi^\dagger(z)e^{-\Omega_\mp(\ta)}$. 
We use the vertex operator expression for $\psi(z)$ and  $\psi^\dagger (z)$
\be\label{psi1}
 \psi(z)=e^{\alpha_0}z^{\alpha_0}e^{-\sum_{i<0} \frac{\alpha_i}{z^i}}e^{-\sum_{i>0}\frac{\alpha_{i}}{z^i}}
\ee
\be\label{psi2}
 \psi^\dagger(z)=e^{-\alpha_0}z^{-\alpha_0}e^{\sum_{i<0} \frac{\alpha_i}{z^i}}e^{\sum_{i>0}\frac{\alpha_i}{z^i}}
\ee
where 
 \be\label{currents}
  \alpha_m=\sum_{i\in\mathbb{Z}}\,:\psi_i\psi^\dag_{i+m}:\,
  \ee
for future use we also introduce
  \be
  \g (\bpow):= e^{\sum_{m=1}^\infty t_m\alpha_m} \,,\quad
  \g^\dag (\bpow):= e^{\sum_{m=1}^\infty t_m\alpha_{-m}} 
  \ee
Note that $[\alpha_i,\alpha_j]=i\delta_{i,-j}$, hence they form a Heisenberg algebra.
Now use the standard realization of the Heisenberg algebra 
\[
\alpha_k=\partial_k,\quad \alpha_{-k}=kt_k, \quad \alpha_0=q\partial_q, \quad e^{\alpha_0}=q
\]
Then (\ref{psi1}), respectively (\ref{psi2}) turn into
\be\label{psi3}
 \psi(z)=qz^{q\partial_q}e^{\sum_{i=1}^\infty t_i z^i}e^{-\sum_{i=1}^\infty \partial_i \frac{z^{-i}}{i}}
\ee
\be\label{psi4}
 \psi^\dagger(z)=q^{-1}z^{-q\partial_q}e^{-\sum_{i=1}^\infty t_i z^i}e^{\sum_{i=1}^\infty \partial_i \frac{z^{-i}}{i}}
\ee
Using the
 following formulas which can easily be deduced from the Cambell-Baker-Hausdorff formula:
\[
 e^{a\partial_x}e^{bx}=e^{bx}e^{a(b+\partial_x)},\qquad e^{a\partial_x^2}e^{bx}=e^{bx}e^{a(b+\partial_x)^2}
\]
one thus obtains:
\be
\label{Psi1}
 \Psi_\mp(z)=(1-z^2)^{\frac12 \pm \frac12}qz^{q\partial_q}
 e^{\sum_{i=1}^\infty t_i z^i}e^{-\sum_{i=1}^\infty \partial_i \frac{z^{-i}+z^{i}}{i}}
\ee
\be
\label{Psi2}
 \Psi_\mp^\dagger(z)=(1-z^2)^{\frac12 \mp \frac12}q^{-1}z^{-q\partial_q}
 e^{-\sum_{i=1}^\infty t_i z^i}e^{\sum_{i=1}^\infty \partial_i \frac{z^{-i}+z^{i}}{i}}
\ee
Hence the orthogonal and symplectic characters satisfy the bilinear equation:
\be 
\label{KP3}
{Res}_z \Psi_\mp(z)\sigma_\mp\otimes \Psi_\mp^\dagger(z)\sigma_\mp=0,
\ee
or equivalently
\be 
\label{KP4}
{Res}_z (1-z^2)e^{\sum_{i=1}^\infty (t_i-s_i)z^i}\sigma_\mp(t-[z]-[z^{-1}]) \sigma_\mp(s+[z]+[z^{-1}])=0,
\ee
which is equation (5.4) of Baker \cite{Baker}.

Now note that if we write $\Psi_\mp(z)=\sum_{i\in\mathbb{Z}}\Psi_{\mp i}z^i$ and 
$\Psi_\mp^\dagger(z)=\sum_{i\in\mathbb{Z}}\Psi^\dagger _{\mp i}z^{-i-1}$, then the modes still satisfy the usual relations.
\[
 \Psi_{\mp i}\Psi_{\mp j}+\Psi_{\mp j}\Psi_{\mp i}=0=\Psi_{\mp i}^\dagger
 \Psi_{\mp j}^\dagger+\Psi_{\mp j}^\dagger\Psi_{\mp i}^\dagger
\qquad \Psi_{\mp i}\Psi_{\mp j}^\dagger+\Psi_{\mp j}^\dagger\Psi_{\mp i}=\delta_{ij}
\]
Using the above vertex operators on the vacuum $|0\rangle=q^0$ one still has
\[
 \Psi_{\mp i}|0\rangle=0=\Psi_{\mp (-i-1)}^\dagger|0\rangle\qquad i<0
\]

Another approach is as follows, equation (\ref{KP3}) still  generates the $GL_\infty$ goup orbit of the vacuum, 
however one has to take a different realization of $gl_\infty$,
 viz. $\Psi_{\mp i}\Psi_{\mp j}^\dagger$  ($i,j\in\mathbb{Z}$) still forms a basis of $gl_\infty$, it is the 
 coefficients of $z^iy^{-j-1}$ in the expansion
\[
 X_\mp(y,z)=e^{\Omega_\mp(\ta)}\psi(z)\psi^\dagger(y)e^{-\Omega_\mp(\ta)}=
\Psi_\mp(z)\Psi_\mp^\dagger(y)
\]
Using the above vertex operators we find
\[
 X_\mp(y,z)=\frac{(1-z^2)^{\frac12 \pm \frac12}(1-y^2)^{\frac12 \mp \frac12}}{(z-y)(1-zy)}\left(z/y\right)^{q\partial_q}
e^{\sum_{i=1}^\infty t_i (z^i-y^i)}
e^{-\sum_{i=1}^\infty \partial_i \frac{z^{-i}+z^{i}-y^{-i}-y^{i}}{i}}
\]

Clearly also the standard Heisenberg algebra changes.

Now define
$\beta^\mp=e^{\Omega_\mp}\alpha_k e^{\Omega_\mp}$, then the $\beta_k$ still have the standard commutation relations, 
$[\beta^\mp_i,\beta^\mp_j]=i\delta_{i,-j}$,
however these elements arealized in a different way.
Using \[
 e^{a\partial_x}x=(x+a)e^{a\partial_x},\qquad e^{a\partial_x^2}x=(x+2a\partial_x)e^{a\partial_x^2}
\]
Hence,
\[
 \beta^\mp_k=\partial_k\quad \beta^\mp_{-k}=kt_k-\partial_k\mp\delta_{k,{\rm even}}, \quad \beta^\mp_0=q\partial_q
\]
And clearly 
\[
 \Psi_\mp(z)=e^{\beta^\mp_0} z^{\beta^\mp_0}e^{-\sum_{i<0} \frac{\beta^\mp_{i}}{i}{z^i}}
 e^{-\sum_{i>0}\frac{\beta^\mp_{i}}{i}{z^i}}
\]
\[
 \Psi_\mp^\dagger(z)=e^{-\beta^\mp_0} z^{-\beta^\mp_0}e^{\sum_{i<0} \frac{\beta^\mp_{i}}{i}{z^i}}
 e^{\sum_{i>0}\frac{\beta^\mp_{i}}{i}{z^i}}
\]
or equivalently
\be
\label{Psi3}
 \Psi_\mp(z)=q z^{q\partial_q}e^{\sum_{i=1}^\infty \left(t_i -\frac{\partial_i\pm \delta_{i,{\rm even}}}{i}\right){z^i}}
 e^{-\sum_{i=1}^\infty \partial_i \frac{z^{-i}}{i}}
\ee
\be
\label{Psi4}
 \Psi_\mp^\dagger(z)=q^{-1} z^{-q\partial_q}
 e^{-\sum_{i=1}^\infty \left(t_i -\frac{\partial_i\pm \delta_{i,{\rm even}}}{i}\right){z^i}}
 e^{\sum_{i=1}^\infty \partial_i \frac{z^{-i}}{i}}
\ee
Note that one obtains (\ref{Psi1}), respectively  (\ref{Psi2}) from (\ref{Psi3}), respectively (\ref{Psi4}), 
if one moves the differential operator part to the right. One can use 
\[
 e^{a\partial_x + bx}=e^{\frac{ab}2}e^{bx}e^{a\partial_x}
\]
Note that 
\[
 X_\mp(y,z)=\frac{1}{(z-y)}\left(z/y\right)^{q\partial_q}
e^{\sum_{i=1}^\infty \left(t_i-\frac{\partial_i\pm \delta_{i,{\rm even}}}{i}\right) (z^i-y^i)}
e^{-\sum_{i=1}^\infty \partial_i \frac{z^{-i}-y^{-i}}{i}}
\]

The above suggest that we can take the normal Clifford algebra in $\psi_i$ and $\psi^\dagger_j$, but choose another 
realization of the Heisenberg algebra, viz., the ones given by the $\beta_i^\mp$, such that the fields $\psi(z)$ and 
$\psi^\dagger(z)$ are given by (\ref{Psi3}), respectively (\ref{Psi4}).
Then the tau function, which is in the KP hierarchy given by
\[
 \tau(\bpow)=\langle 0| e^{\sum_{i>1} t_i\alpha_i}g|0\rangle\qquad\mbox{for } g\in Gl_\infty
\]
changes into $\sigma_\mp(\bpow)=e^{\Omega_\mp}\tau(\bpow)$, which is equal to
\[
\sigma_\mp(\bpow)= \langle 0|e^{\sum_{i>1} t_i\alpha_i-\frac{1}{2i}\alpha_i^2\mp 
\frac{1}{2i}\alpha_{2i}}g|0\rangle\qquad\mbox{for } g\in Gl_\infty
\]
which corresponds to the modified Hamiltonian of \cite{Baker}, Section 3, Approach I.
Next calculate 
\[
\langle 0|e^{\sum_{i>1} t_i\alpha_i-\frac{1}{2i}\alpha_i^2\mp \frac{1}{2i}\alpha_{2i}}
e^{\sum_{i>1} t_i^*\alpha_{-i}}|0\rangle=\tau_\mp(\bpow^*|\bpow)
\]
Hence, it makes sense to look at 
\[
\langle 0|e^{\sum_{i>1} t_i\alpha_i-\frac{1}{2i}\alpha_i^2\mp \frac{1}{2i}\alpha_{2i}}g
e^{\sum_{i>1} t_i^*\alpha_{-i}}|0\rangle\qquad\mbox{for } g\in Gl_\infty
\]

\br

Actually we have
\be
\tau (\bpow,\bpow^*) \to
{\tau}^{\pm}(\bpow,\bpow^*)=e^{\Omega_\pm({\tilde\partial})}\cdot \tau (\bpow,\bpow^*)
\ee
If
\[
 \tau (\bpow,\bpow^*)=\sum_{\lambda,\mu\in\Pa} s_\lambda(\bpow) \pi_{\lambda,\mu} s_\mu(\bpow^*) 
\]
where
\[
\pi_{\lambda,\mu}=\l 0|\,s_\lambda({\tilde\alpha})\,g\, s_\mu({\tilde\alpha}^*)\,|0\r
\]
then
\[
 \tau^\pm (\bpow,\bpow^*)=\sum_{\lambda,\mu\in\Pa} s_\lambda(\bpow) \pi_{\lambda,\mu}^\pm s_\mu(\bpow^*)\,,\quad 
\]
where
\[ 
\pi_{\lambda,\mu}^+=\l 0|\,sp_\lambda({\tilde\alpha})\,g\, s_\mu({\tilde\alpha}^*)\,|0\r \,,\quad
\pi_{\lambda,\mu}^-=\l 0|\,o_\lambda({\tilde\alpha})\,g\, s_\mu({\tilde\alpha}^*)\,|0\r
\]
Similarly, one can consider $\tau^{a,b}$ with $a,b = \pm$.

\er

 \section{Integrals over symplectic group and over orthogonal groups}

\paragraph{Haar measures and generating functions for characters.}

Lemma \ref{Vand(z)} in Appendix \ref{Rewriting Vandermond} and formulae of the Appendix \ref{Haar-integration} results in
the following lemmas we shall need:
\bl\label{lemma-Haar-Sp-U} {\em 
The Haar measures of the symplectic group $\mathbb{S}p(2n)$ and of the unitary group 
 $\mathbb{U}(n)$ are related as follows
 \bea\label{Haar-Sp-U-f--}
e^{\tr V\left( S,\bpow \right)}d_*S &=& 2^{-n} \tau_-(U|\bpow) \tau_-(U^{-1}|\bpow)d_*U
 \\
 \label{Haar-Sp-U-f++}
&=& 2^{-n}  \tau_+(U|\bpow) \tau_+(U^{-1}|\bpow)\det(1-U^2)\det(1-U^{-2})d_*U
 \\
 \label{Haar-Sp-U-f+-}
&=& 2^{-n}  \tau_+(U|\bpow) \tau_-(U^{-1}|\bpow)\det(1-U^2)d_*U
 \eea 
 where $e^{i\theta_1},e^{-i\theta_1},\dots, e^{i\theta_n},e^{-i\theta_n}$ are eigenvalues of $S\in \mathbb{S}p(2n)$ 
 while $e^{i\theta_1},\dots, e^{i\theta_n}$
 are eigenvalues of $U\in \mathbb{U}(n)$. 
}
 \el

 \begin{Lemma} The Haar measures of the orthogonal group 
 $\mathbb{O}(2n)$ and of the unitary group $\mathbb{U}(n)$
 are related as follows
 \bea\label{Haar-O-even-U-f++}
e^{\tr V( O,\bpow)}d_*O &=&  2^{-n}  \tau_+(U|\bpow) \tau_+(U^{-1}|\bpow)d_*U
 \\
 \label{Haar-O-even-U-f--}
& = & 2^{-n}  \tau_-(U|\bpow) \tau_-(U^{-1}|\bpow)\det(1-U^2)^{-1}\det(1-U^{-2})^{-1}d_*U
 \\
 \label{Haar-O-even-U-f-+}
& = & 2^{-n} \tau_-(U|\bpow) \tau_+(U^{-1}|\bpow)\det(1-U^2)^{-1}d_*U
\eea
\end{Lemma}

\begin{Lemma}
 The Haar measures of the orthogonal group $\mathbb{O}(2n+1)$ and of the unitary group $\mathbb{U}(n)$
 are related as follows
 \bea\label{Haar-O-odd-U-f++}
e^{\tr V( O,\bpow)}d_*O &=&  2^{-n}  \tau_+(U|\bpow) \tau_+(U^{-1}|\bpow)\det(1-U)\det(1-U^{-1})d_*U 
\\
\label{Haar-O-odd-U-f--}
& = & 2^{-n}  \tau_-(U|\bpow) \tau_-(U^{-1}|\bpow)\det(1+U)^{-1}\det(1+U^{-1})^{-1}d_*U
\\
\label{Haar-O-odd-U-f-+}
& = & 2^{-n}  \tau_-(U|\bpow) \tau_+(U^{-1}|\bpow)\det\frac{1-U^{-1}}{1+U}d_*U
 \eea
 where $e^{i\theta_1},e^{-i\theta_1},\dots, e^{i\theta_n},e^{-i\theta_n},1$ are eigenvalues of $O\in \mathbb{O}(2n+1)$ 
 while $e^{i\theta_1},\dots, e^{i\theta_n}$
 are eigenvalues of $U\in \mathbb{U}(n)$. 
 \end{Lemma}

  \subsection{Integrals over symplectic group.}
  
  Consider the following integral over the symplectic group
 \be\label{Haar-symplectic}
I_{\mathbb{S}p(2n)}(\bpow)=\int_{S\in\mathbb{S}p(2n)} e^{\sum_{m=1}^\infty t_m\tr S^m}d_*S
 \ee
 where $d_*S$ is the corresponding Haar measure. Explicitly 
 \be\label{eigen-symplectic}
I_{\mathbb{S}p(2n)}(\bpow)=\frac{2^{n^2}}{\pi^n } 
\int_{0\le \theta_1 \le \cdots \le \theta_n \le \pi}\,\prod_{i<j}^n\,(\cos \theta_i -\cos \theta_j )^2 \,
\prod_{i=1}^n e^{2\sum_{m=1}^\infty t_m \cos m\theta_i} \sin^2 {\theta_i} d\theta_i
 \ee
where $e^{\pm i\theta_1},\dots,e^{\pm i\theta_n}$ are the eigenvalues of $S$.

By analogy with matrix models studied in physics we call the parameters $\bpow=(p_1,p_2,\dots)$ coupling constants, and 
(asymptotic) series in these parameters are called perturbation series.

\paragraph{Perturbation series for integrals over symplectic group as series in characters $o_\lambda(\bpow)$, 
$sp_\lambda(\bpow)$, $s_\lambda(\bpow)$.}

\bp {\em
\bea
\label{symplectic-int-via-Schurs}
I_{\mathbb{S}p(2n)}(\bpow)
&=& \sum_{\lambda\in\mathbb{P}\atop \ell(\lambda)\le n} s_{\lambda\cup \lambda}(\bpow)
\\
\label{symplectic-int-via-oo}
&=& 2^{-n} \sum_{\lambda\in\mathbb{P}\atop \ell(\lambda)\le n} \left(o_\lambda(\bpow)\right)^2=
2^{-n} \sum_{\lambda\in\mathbb{P}\atop \lambda_1\le n} \left(sp_\lambda(-\bpow)\right)^2
\eea
}
\ep

 Formula (\ref{symplectic-int-via-Schurs}) may be derived using of Cauchy-Littlewood formula
 \[
 e^{\sum_{m=1}^\infty \frac 1m p_m\tr S^m} =\sum_\lambda s_\lambda(S)s_\lambda(\bpow)
 \]
 and the known relation (see for instance (6.13)-(6.15) in Sect 6, VII in \cite{Mac})
 \be\label{Schur-over-Sp-group} 
\int_{S\in \mathbb{S}p(2n)} s_\lambda(S) d_*S =
\cases{
1 \quad \lambda^{tr}\,\mbox{is\,even}  \cr
0 \quad\, \mbox{otherwise}
 }
  \ee  
where $\lambda^{tr}$ is the partition conjugated to $\lambda$, see \cite{Mac}. (This relation may be easily obtained
 by the evaluation of the Schur function 
$s_\lambda(z)$ where  $z_i=x_i+x_i^{-1}$ inside the integral over symplectic group).
 
The left hand side of (\ref{symplectic-int-via-oo}) is obtained from (\ref{Haar-Sp-U-f--}), (\ref{f-t*}) 
and (\ref{orthonormality-ch-U}), and the right hand side of (\ref{symplectic-int-via-oo}) is the result of (\ref{duality}).

\paragraph{Integrals over symplectic group as DKP tau functions.} Here we need the fermionic language of Section 
\ref{characters-and-fermions}, see also  Appendix \ref{Fermions}.
 \bp\label{symplectic-in-fermions}{\em
 \bea
I_{\mathbb{S}p(2n)}(\bpow)\,=\, \label{symplectic-in-fermi-fields}
\frac{1}{n!}\,\l 2n |\,\g(\bpow)\,  e^{\frac{1}{4\pi i}\oint \psi(x^{-1})\psi(x)(x-x^{-1})\frac{dx}{x}}\, |0\r 
 \\ 
=\,\frac{1}{n!}\,\l 2n |\,\g(\bpow)\,   \label{symplectic-in-fermi-modes}
e^{\sum_{i\in\mathbb{Z}}\,\psi_i\psi_{i-1}}\, |0\r
 \eea
 }
 \ep

{\bf Proof.} The second equality follows from
    \be\label{int=sum-symplectic}
 \frac{1}{4\pi i}\oint \psi(x^{-1})\psi(x)(x-x^{-1})\frac{dx}{x} = \sum_{i\in\mathbb{Z}} \psi_i\psi_{i-1}   
    \ee

    Let us consider the Taylor series of the exponential in the right hand side.    
    The first equality follows from 
    \be
  \l N|\psi(x_1)\cdots \psi(x_N) |0\r=\prod_{i<j} (x_i-x_j)=:\Delta_N(x)  
    \ee  
    and 
    \be\label{Delta-exp=Delta-cos}
 \Delta_{2n}(e^{-i\theta_1},e^{i\theta_1},\dots,e^{-i\theta_n},e^{i\theta_n})  =
 (-i)^n 2^{n^2}\prod_{k<j}(\cos \theta_k -\cos \theta_j)^2\prod_{k=1}^n \sin {\theta_k} 
  \ee

Let us mark that formula (\ref{symplectic-int-via-Schurs}) follows from (\ref{int=sum-symplectic}) and from results of \cite{OST-I} 
   (see formulae for $S^{(1)}_{4}$ in \cite{OST-I}).

 \paragraph{Pfaffian representation.} Here we need Appendix \ref{Pfaffians}.
Let us note that thank to the Wick's rule we directly obtain the Pfaffian representation of the integral 
(\ref{Haar-symplectic}) as follows
 \bp {\em
 \be\label{pfaffian-rep-symplectic}
I_{\mathbb{S}p(2n)}(\bpow)= \Pf \left[M_{kj}(\bpow)\right]_{k,j=1,\dots,2n}
 \ee
 where $M$ is the following Toeplitz matrix
 \be\label{M-symplectic}
 M_{kj}(\bpow)=-M_{jk}(\bpow)=
 \frac{1}{4\pi i}\oint \left(x^{j-k}-x^{k-j} \right)(x-x^{-1})e^{\sum_{m=1}^\infty \frac 1m p_m(x^m+x^{-m}) }\frac{dx}{x}
 \ee
 }
 \ep

\paragraph{Relation of the integral over $\mathbb{S}p(2n)$ to an integral over $\mathbb{U}(2n)$.} 
We can either refer to results of \cite{LeurO} about the interplay between BKP and two-component KP tau functions, or
do the following.  Consider 

 \be
I_{\mathbb{U}(2n)}(\bpow):= \int_{U\in\mathbb{U}(2n)} \det\left(U-U^\dag \right) 
   e^{\sum_{m=1}^\infty \frac 1m p_m \left(\tr U^m + \tr U^{-m}\right)} d_*U
\ee
Written as an integral over eigenvalues it is
 \be
I_{\mathbb{U}(2n)}(\bpow) = \frac{1}{(2n)!(2\pi)^{2n}}\oint \Delta_{2n}(x)\Delta_{2n}(x^{-1}) \prod_{i=1}^{2n} (x_i-x_i^{-1})
 e^{\sum_{m=1}^\infty \frac 1m p_m \left(x_i^m + x_i^{-m}\right)} \frac{dx_i}{x_i} 
 \ee

We have
\begin{Lemma}\label{lemma-vev-for-U(2n)}
 \be\label{vev-for-U(2n)}
 I_{\mathbb{U}(2n)}(\bpow)= 
 \l 2n,-2n|e^{\sum_{m=1}^\infty \frac 1m p_m\left(\alpha_m^{(1)}-\alpha_m^{(2)}\right)} e^{\frac{1}{2\pi i}
 \oint \psi^{(1)}(x^{-1}) \psi^{\dag(2)}(x) (x-x^{-1})\frac{dx}{x} }\,|0,0\r
 \ee
 For the two-component fermions see Appendix, $\alpha^{(a)}_n:=\sum_{i\in\mathbb{Z}}\psi^{(a)}_i\psi^{\dag(a)}_{i+n}
 ,\, n>0$. 
\end{Lemma}
The Lemma is the direct result of \cite{HO-2006} (see Proposition 4 there). We also have
    \bp\label{Prop-U(2n)=Sp(2n)}{\em
    \be
    \left(2^n\int_{S\in\mathbb{S}p(2n)} e^{\sum_{m=1}^\infty \frac 1m p_m\tr S^m}d_*S  \right)^2 = 
    \int_{U\in\mathbb{U}(2n)} \det\left(U-U^\dag \right) 
    e^{\sum_{m=1}^\infty \frac 1m p_m \left(\tr U^m + \tr U^{-m}\right)} d_*U
    \ee
    }
   \ep  
   Proof. One way to prove it is to present $I_{\mathbb{U}(2n)}$ as determinant of $M$ of (\ref{M-symplectic}).
  It is easy using the Wick's rule for the vacuum expectation value in Lemma \ref{lemma-vev-for-U(2n)}. Via Wick's rule we directly 
  obtain
 \be
 I_{\mathbb{U}(2n)}(\bpow)=\det \left[ 2M_{ij}  \right]_{i,j=1,\dots,2n}
 \ee
 with the same matrix $M$ as in Pfaffian representation (\ref{pfaffian-rep-symplectic}). This proves (\ref{Prop-U(2n)=Sp(2n)}).
  
 The other way is to apply the results of \cite{LeurO} (see Proposition 4 there). Then we have
  \be
 \left( 2^n I_{\mathbb{S}p(2n)}(\bpow)\right)^2 =
 \l 2n,-2n|e^{\sum_{m=1}^\infty \frac 1m p_m\left(\alpha_m^{(1)}-\alpha_m^{(1)}\right)} e^{\frac{1}{4\pi i} \oint \psi^{(1)}(x^{-1})
 \psi^{\dag(2)}(x) (x-x^{-1})\frac{dx}{x} }\,|0,0\r
  \ee
which coincides with the right hand side of (\ref{vev-for-U(2n)}). The end of proofs.

\paragraph{Different fermionic representation and Toda chain-AKNS tau function}

Denote $z=x+x^{-1}$. Introduce variables ${\tilde \bpow}=({\tilde p}_1,{\tilde p}_2,\dots)$ with the help
 \be\label{p-tilde{p}}
\sum_{n=1}^\infty \frac 1n p_n(x^n+x^{-n})\,=\,\sum_{n=1}^\infty \frac 1n {\tilde p}_n z^n -c({\tilde \bpow})
,\quad c({\tilde\bpow})=\sum_{n} \frac{(2n)!}{(n!)^2}\frac{{\tilde p}_{2n}}{2n}
 \ee
 where the sets of $\{{\tilde p}_{2n},\, n > 0  \}$ and of $\{{\tilde p}_{2n+1},\,n\ge 0  \}$ may be expressed via
 the sets
  $\{{ p}_{2n},\, n > 0  \}$ and of $\{{ p}_{2n+1},\,n\ge 0  \}$ respectively by triangle transformations with binomial 
 entries.

Then thanks to (\ref{l1}), (\ref{l4}) in Appendix \ref{Rewriting Vandermond} we have
\be\label{I_Sp=1MM}
I_{\mathbb{S}p(2n)}(\bpow({\tilde \bpow }))=\,\int_{-2}^{2}\cdots  \int_{-2}^{2} \prod_{i<j} (z_i-z_j)^2 
\prod_{i=1}^n e^{\sum_{m=1}^\infty \frac 1m \tilde{p}_m z_i^m} (4-z_i^2) dz_i
\ee
The last integral is an example of well-studied $\beta=2$ ensemble and may be presented in form of a determinant:
 \be
 I_{\mathbb{S}p(2n)}(\bpow({\tilde \bpow }))=\det [N_{ij}({\tilde \bpow })]_{i,j=1,\dots,n}
 \ee
 where $N_{ij}$ are the so-called moments. In our case
 \be
 N_{ij}({\tilde \bpow })=\int_{-2}^{2} z^{i+j} e^{\sum_{m=1}^\infty \frac 1m \tilde{p}_m z^m}(4-z^2) dz
 \ee
The fermionic representation for $\beta=2$ ensembles is known, see  , in our case
\[
 =
\]

\subsection{Integrals over orthogonal group.}

Consider the following integral over the orthogonal group $\mathbb{O}(2n)$
 \be\label{Haar-orthogonal-even}
I_{\mathbb{O}(N)}(\bpow)=\int_{O\in\mathbb{O}(2n)} e^{\sum_{m=1}^\infty \frac1m p_m\tr O^m}d_*O
 \ee
 where $d_*O$ is the corresponding Haar measure. Explicitly 
 \be\label{eigen-orthogonal-even}
I_{\mathbb{O}(2n)}(\bpow)=\frac{2^{(n-1)^2}}{\pi^n n!} 
\int_{0\le \theta_1 \le \cdots \le \theta_n \le \pi}\,\prod_{i<j}^n\,(\cos \theta_i -\cos \theta_j )^2 \,
\prod_{i=1}^n e^{2\sum_{m=1}^\infty \frac 1m p_m \cos m\theta_i}  d\theta_i
 \ee
where $e^{\pm i\theta_1},\dots,e^{\pm i\theta_n}$ are the eigenvalues of $O$.
and
 \be\label{eigen-odd-othogonal}
I_{\mathbb{O}(2n+1)}(\bpow)=\frac{2^{n^2}}{\pi^n n!} 
\int_{\theta_1 \le \cdots \le \theta_n \le \pi}\,\prod_{i<j}^n\,(\cos \theta_i -\cos \theta_j )^2 \,
\prod_{i=1}^n \sin^2\frac{\theta_i}{2} e^{\sum_{m=1}^\infty \frac 1m p_m (1+2\cos m\theta_i)} d\theta_i
 \ee
where $e^{\theta_1},e^{-\theta_1},\dots,e^{\theta_n},e^{-\theta_n},1$ are eigenvalues of $O(2n+1)$. 
 
 \br \label{Lemma-O-from-Sp}
 Then
  \be\label{symplectic-othogonal-via-vertexB}
 I_{\mathbb{O}(2n)}(\bpow)\,=\,2I_{\mathbb{S}p(2n)}(\bpow+[1]+[-1])
 \ee
  \be\label{eigen-odd-othogonal-via-vertexB}
 I_{\mathbb{O}(2n+1)}(\bpow)= \,\frac 12 \, B^+(1)I_{\mathbb{O}(2n)}(\bpow) := \,
 \frac 12 \,e^{\sum_{m=1}^\infty \frac 1m p_m } \, I_{\mathbb{O}(2n)}(\bpow-[1])
 \ee
 \er

According to (\ref{Haar-O-even-U-f++}) and 
according to (\ref{f-t*}), (\ref{f+t*}) we obtain
\be\label{int-dU=int-dO-even-f++}
I_{\mathbb{O}(2n)}(\bpow)=2^{1-n}\int_{\mathbb{U}(n)} \tau_+(U|\bpow)\tau_+(U^{-1}|\bpow)d_*U
\ee
\be\label{int-dU=int-dO-even-f--}
I_{\mathbb{O}(2n)}(\bpow)=2^{1-n}\int_{\mathbb{U}(n)} \tau_-(U|\bpow)\tau_-(U^{-1}|\bpow)\det(U-U^\dag)^{-2}d_*U
\ee

\paragraph{Perturbation series for integrals over orthogonal group as series in characters $sp_\lambda(\bpow)$, 
$o_\lambda(\bpow)$, $s_\lambda(\bpow)$.}

\bp {\em
\bea
\label{orthogonal-Schurs}
I_{\mathbb{O}p(2n)}(\bpow)
&=& 
   I_{\mathbb{O}(N)}(\bpow)=\sum_{\lambda {\rm even}\atop \ell(\mu)\le N} s_{\lambda }(\bpow)  
\\
\label{orthogonal-int-via-spsp}
&=& 2^{1-n} \sum_{\lambda\in\mathbb{P}\atop \ell(\lambda)\le n} \left(sp_\lambda(\bpow)\right)^2 =
2^{1-n} \sum_{\lambda\in\mathbb{P}\atop \lambda_1 \le n} \left(o_\lambda(-\bpow)\right)^2
\eea
}
\ep

Equality (\ref{orthogonal-Schurs}) follows from
the Cauchy-Littlewood formula
 \[
 e^{\sum_{m=1}^\infty \frac 1m p_m\tr O^m} =\sum_\lambda s_\lambda(O)s_\lambda(\bpow)
 \]
and (see for instance (3.19)-(3.21) in Sect 3, VII in \cite{Mac})
 \be\label{} 
\int_{O\in \mathbb{O}(N)} s_\lambda(O) d_*O =
\cases{
1 \quad \lambda \,\mbox{is\,even}  \cr
0 \quad\, \mbox{otherwise}
 }
  \ee  
 
The left hand side of (\ref{orthogonal-int-via-spsp}) follows from (\ref{Haar-O-even-U-f++}) , (\ref{f+t*})) 
and (\ref{orthonormality-ch-U}), then the right hand side is the result of (\ref{duality}).

\paragraph{Integrals over orthogonal group as BKP tau functions.}

 \bp \label{orthogonal-in-fermions}{\em
 \bea
n! \,I_{\mathbb{O}(2n)}(\bpow)\,
=\,\l 2n |\,\g(\bpow)\, \label{orthogonal-even-in-fermi-fields}
e^{\frac{1}{4\pi i}\oint \psi(x^{-1})\psi(x)(x^{-1}-x)^{-1}\frac{dx}{x}}\, |0\r 
 \\
=\,\l 2n | \,\g(\bpow)\, \label{orthogonal-even-in-fermi-modes}
e^{\sum_{k \ge 0}\sum_{i\in\mathbb{Z}}\,\psi_i\psi_{i-1-2k}} |0\r
 \eea

 \bea
n! \,I_{\mathbb{O}(2n+1)}(\bpow)\,=\, \label{orthogonal-odd-in-fermi-fields}
e^{\sum_{m>0} \frac 1m p_m}\l 2n+1 |\,\g(\bpow)\,
e^{\frac{1}{4\pi i}\oint \psi(x^{-1})\psi(x)\frac{x-1}{x+1}\frac{dx}{x} }\,|0\r 
 \\
=\, e^{\sum_{m>0} \frac 1m p_m}\l 2n+1 |\,\g(\bpow)\, \label{orthogonal-odd-in-fermi-modes}
e^{\sum_{i > j}\,(-)^{i+j}\psi_i\psi_{j}}\, |0\r
  \eea
  or, it can be written in an unified way as
  \bea
n! \,I_{\mathbb{O}(N)}(\bpow)\,=\,  \label{orthogonal-via-Fermi-fields}
\l N | \,\g(\bpow)\,
e^{\frac{1}{4\pi i}\oint \psi(x^{-1})\psi(x) (x^{-1}-x)^{-1}   \frac{dx}{x} + \frac{1}{\sqrt 2}\phi \psi(1)} \,|0\r 
 \\
=\,\l N |\,\g(\bpow)\, \label{orthogonal-via-Fermi-modes}
e^{-\sum_{k\ge 0}\sum_{i\in\mathbb{Z}}\,\psi_i\psi_{i-1-2k} + \frac{1}{\sqrt 2}\phi\sum_{i\in\mathbb{Z}} \psi_{i}}\, |0\r
  \eea
 where $N$ may be even or odd.}
 \ep
 
 The simplest way to prove this Proposition is just to apply Lemma \ref{Lemma-O-from-Sp} to Proposition 
 \ref{symplectic-in-fermions}. However let us prove it directly from the fermionic expressions like we did
 in the proof of Proposition \ref{symplectic-in-fermions}.
 
 It is convenient to re-write the fermionic exponent in (\ref{orthogonal-even-in-fermi-fields}):
    \be\label{int-in-theta-orthogonal}
     \frac{1}{4\pi i}\oint \psi(x^{-1})\psi(x)(x^{-1}-x)\frac{dx}{x}=
     \frac{i}{\pi} \int_0^\pi \frac{\psi(e^{-i\theta})\psi(e^{i\theta})}{\sin \theta} d\theta
    \ee
    where we note that $\frac{\psi(e^{-i\theta})\psi(e^{i\theta})}{\sin \theta}=i(\psi(1)\psi'(1)-\psi'(1)\psi(1)$ 
    is not singular at $\theta=0$. The proof basically repeats the the proof of Proposition \ref{symplectic-in-fermions}. 
    For the vacuum expectation value in the right hand side of (\ref{orthogonal-even-in-fermi-fields})  we obtain  
     \be\label{vev=integral-in-theta-orthogonal-even}
 \frac{1}{n!}\, \left( \frac{i^n}{\pi^n n!} \int_0^\pi \cdots \int_0^\pi 
  \Delta_{2n}(e^{-i\theta_1},e^{i\theta_1},\dots,e^{-i\theta_n},e^{i\theta_n}) 
  \prod_{i=1}^{n} e^{\sum_{m=1}^\infty \frac 1m p_m\cos m\theta_i} \frac{d\theta_i }{\sin \theta_i}\right)
  \ee
 Then formula (\ref{Delta-exp=Delta-cos}) proves (\ref{orthogonal-even-in-fermi-fields}). The formula
 (\ref{orthogonal-even-in-fermi-modes}) then follows from
  \[
  \frac{1}{4\pi i}\oint \psi(x^{-1})\psi(x)(x^{-1}-x)^{-1}\frac{dx}{x} =
  \frac{1}{4\pi i}\oint \psi(x^{-1})\psi(x) \frac{dx}{1-x^2} = 
  \frac 12 \sum_{k \ge 0}\sum_{i\in\mathbb{Z}}\,\psi_i\psi_{i-1-2k}
  \]
Formula (\ref{orthogonal-odd-in-fermi-fields}) and the unifying Formula (\ref{orthogonal-via-Fermi-fields}) follows from 
(\ref{eigen-odd-othogonal-via-vertexB}) and the bosonization relations, see Appendix \ref{Vertex operators}. At last
\[
 \frac{1}{4\pi i}\oint \psi(x^{-1})\psi(x)\frac{x-1}{x+1}\frac{dx}{x}=
 \frac{1}{2\pi i}\oint \psi(x^{-1})\psi(x)(x-x^2+x^3-\cdots)\frac{dx}{x}
 =\sum_{i > j}\,(-)^{i+j}\psi_i\psi_{j}
\]
Thus each relation of Proposition \ref{orthogonal-in-fermions} is proven.

\paragraph{Pfaffian representation.} 
 
Let us mark that thanks to the Wick's rule applied to calculate (\ref{orthogonal-even-in-fermi-fields}) we directly 
obtain the Pfaffian representation of the integral (\ref{eigen-orthogonal-even}) as follows
 \bp {\em
 \be\label{pfaffian-rep-orthogonal}
I_{\mathbb{O}(2n)}(\bpow)= \Pf \left[M_{kj}(\bpow)\right]_{k,j=1,\dots,2n+2}
 \ee
 where in case $N$ even
 \be\label{M-orthogonal}
 M_{kj}(\bpow)=\frac{1}{4\pi i}\oint \left(x^{j-k}-x^{k-j} \right)(x-x^{-1})^{-1}
 e^{\sum_{m=1}^\infty \frac 1m p_m(x^m+x^{-m}) }\frac{dx}{x}
 \ee
 In case $N=2n+1$ the entries $M_{kj}$ $k,j\le 2n+1 $ as before and
  \be
M_{k,2n+2}=-M_{2n+2,k}=  \frac {1}{2\pi i}  \oint x^k e^{\sum_{m=1}^\infty \frac 1m p_m(x^m+x^{-m}) } \frac {dx}{x}
  \ee
  }
 \ep
 For the proof we notice that from Wick's rule we get
\[
 M_{kj}(\bpow)= \frac{1}{2\pi i}
 \l 0|\psi^\dag_k\psi^\dag_j 
 \oint \frac{\psi(x^{-1})\psi(x)}{x^{-1}-x} \frac{dx}{x}|0\r
 \]

\paragraph{Relation of the integral over $\mathbb{O}(N)$ to an integral over $\mathbb{U}(2n)$.}

Next turn to its two-component KP counterpart. According to Proposition 4 in \cite{LeurO} we have
 \[
 \left( I_{\mathbb{O}(N)}(\bpow) \right)^2 =
 \]
  \be
 \l N,-N|e^{\sum_{m=1}^\infty \frac 1m p_m\left(\alpha_m^{(1)}-\alpha_m^{(2)}\right)} 
 e^{\frac{1}{4\pi i} \oint \psi^{(1)}(x^{-1}) \psi^{\dag(2)}(x)
 (x-x^{-1})^{-1}\frac{dx}{x} + 
 \frac{1}{\sqrt 2}\left( \psi^{(1)}(1)\psi^{\dag} - \psi\psi^{\dag(2)}(1) \right)}
 \,|0,0\r
 \ee
 where we use the additional fermions $\psi,\, \psi^\dag$ which anticommute with $\psi^{(i)}(x)$ and $\psi^{\dag(i)}(x)$,
 also $\psi\psi^\dag +\psi^\dag\psi=1$ and $\psi^\dag|*,*\rangle =0$, see Section 2 in \cite{LeurO}. Other
 the notations are the same as in Lemma \ref{lemma-vev-for-U(2n)}.

\paragraph{As 1D TL-NLS tau function.}

Denote $z=x+x^{-1}$ and introduce variables ${\tilde \bpow}=({\tilde p}_1,{\tilde p}_2,\dots)$ with the help
of (\ref{p-tilde{p}}).
Then  similar to (\ref{I_Sp=1MM}) we get
\be
I_{\mathbb{O}(2n)}(\bpow({\tilde \bpow}))=\,\int_{-2}^{2}\cdots  \int_{-2}^{2} \prod_{i<j} (z_i-z_j)^2 
\prod_{i=1}^n e^{\sum_{m=1}^\infty \frac 1m {\tilde p}_m z_i^m - c({\tilde \bpow}) } dz_i
\ee
\be
I_{\mathbb{O}(2n+1)}(\bpow({\tilde \bpow}))=\,\int_{-2}^{2}\cdots  \int_{-2}^{2} \prod_{i<j} (z_i-z_j)^2 
\prod_{i=1}^n e^{\sum_{m=1}^\infty \frac 1m {\tilde p}_m z_i^m -c({\tilde \bpow})} (2-z_i)dz_i
\ee

\subsection{On some integrals over unitary matrices.}
(1) Here we consider a Cauchy-like integral:
\bea\label{U-Cauchy-integral=exp}
 (-1)^{\frac 12 (N^2-N) }
\int_{\mathbb{U}(N)} e^{\sum_{n>0} \frac 1n p_n\tr U^{n}} \det(1-U^\dag)^{-1}\,(\det U)^{N-1} \,d_*U &=&(-1)^{\frac 12 (N^2-N) }
\exp\,\sum_{n>0}\,\frac{p_n}{n} \qquad\qquad
  \\
 = \sum_{\lambda\atop\ell(\lambda)\le N} sp_\lambda(\bpow)
 & =& \sum_{\lambda\atop\ \lambda_1\le N} (-1)^{|\lambda|}o_\lambda(-\bpow)
  \label{sum-of-characters} 
\eea
where we imply $\det(1-U^\dag)^{-1}=\sum_\lambda s_\lambda(U^\dag)s_\lambda(\bpow_o)
,\, \bpow_o=(1,1,\dots) $
(Cauchy-Littlewood identity). 

To prove (\ref{U-Cauchy-integral=exp})
we need
\be\label{scalar-product}
 (2\pi i)^{-N}\oint \det \left( z_i^{h_j}\right)_{i,j} \det \left(z_i^{-{\tilde h}_j}\right)_{i,j} \prod_{i}^N z_i^{L}\frac{dz_i}{z_i}=
 \prod_{j}\delta_{h_j,{\tilde h}_j-L}\, \to \, 
 \int_{\mathbb{U}(N)}s_\lambda(U)s_{\tilde\lambda}(U^\dag)\det U^L d_*U=\delta_{\lambda+L,{\tilde\lambda}}
\ee
which generalizes (\ref{orthonormality-ch-U}). Let us substitute Cauchy-Littlewood series for the exponential and for the
determinant $\det(1-U^\dag)^{-1}$ inside the integral. These seies are respectively $\sum_\lambda s_\lambda(\bpow)s_\lambda(U)$
and $\sum_\lambda s_\lambda(U^\dag)s_\lambda(\bpow_o) $. Using (\ref{scalar-product}) where $L=N-1$, up to the sign factor 
we obtain $\sum_\lambda s_\lambda(\bpow)s_{\tilde\lambda(\lambda)}(\bpow_o)$ where ${\tilde\lambda}_i(\lambda) =\lambda_i+N-1$.
Notice that each $s_\lambda(\bpow_o)=1$ for $\lambda=(n),\, n=0,1,\dots$ and vanish otherwise. Then the first equality is true,
because $\sum_{n\ge 0} s_{(n)}(\bpow)=\exp \sum_{n>0} \frac{p_n}{n}$.

The second equality may be obtained with the help of the Schur-Littlewood relation 
(see Ch. I, Sect. 5, Ex. 4 in \cite{Mac} page 76)
\[
 \sum_{\lambda}\,  s_\lambda(U) = \prod_{i< j}(1-z_i z_j)^{-1}\prod_k(1-z_k)^{-1}
\]
where $z_i,\, i=1,\dots,N$ are eigenvalues of $U$,
written in form
\[
\prod_k(1-z_k^{-1}) \prod_{i< j}(1-z_i^{-1} z_j^{-1})\sum_{\lambda}\,  s_\lambda(U^\dag) =1
\]
and  further re-written in form
\[
 (-1)^{\frac 12 (N^2-N) }(\det U)^{1-N} \det\left(1- U^\dag\right) \,\left(
 \prod_{i< j}(1-z_i z_j)\sum_{\lambda}  \,  s_\lambda(U^\dag)\right) =1
\]
where $z_i,\,i=1,\dots,N$ are eigenvalues of $U$.
Then, inserting the left hand side inside the integral and using first (\ref{f+t*}) and then (\ref{orthonormality-ch-U})  
we obtain (\ref{sum-of-characters}). The third equality follows from (\ref{duality}).
 Relation (\ref{U-Cauchy-integral=exp}) proves that sums of characters
(\ref{sum-of-characters}) are the elementary KP tau function with $\bpow$ being the KP higher times.

(2) Similarly for $N>1$
\bea\label{integral=0}
0=
 (-1)^{\frac 12 (N^2-N) }
\int_{\mathbb{U}(N)} e^{\sum_{n>0} \frac 1m p_n\tr U^{n}}\det \left(1-U^{2} \right)(\det U)^{N-1}  d_*U =
\\ \qquad
\sum_{\lambda\atop\ell(\lambda\cup\lambda)\le N} o_{\lambda\cup\lambda}(\bpow)=
\sum_{\lambda {\rm even}\atop\lambda_1 \le N} sp_{\lambda}(-\bpow)
\label{sums-of-fat-partitions}
\eea

Proof. 
To get the second equality we use 
 \[
 \sum_{\lambda^{\rm tr}\,{\rm even}}\,  s_\lambda(U) = \prod_{i< j}(1-z_i z_j)^{-1}
\]
(see Ch I, Sect 5, Ex 5(b) in \cite{Mac} page 77)
witten in form
\be
 (-1)^{\frac 12 (N^2-N) }\det U^{1-N}\det(1-U^2)^{-1}  \,
 \left(\prod_{i\le j}(1-z_iz_j)\sum_{\lambda}\, s_{\lambda\cup\lambda}(U^\dag)\right) =1
\ee
and relations (\ref{f-t*}) and (\ref{orthonormality-ch-U}). The third equality results from (\ref{duality}). The first
equality in (\ref{integral=0}) is obtained  using 
$\det(1-U^{2})=\sum_\lambda s_\lambda(U) s_{\lambda}(\bpow_o'),\, \bpow_o'=(0,1,0,1,\dots)$ and (\ref{scalar-product}) which gives 0 for each $N$.

\section{The character expansion for $\beta=2$ ensembles}

\paragraph{The character expansions.}

Lemma \ref{Vand(z)} in Appendix  \ref{Rewriting Vandermond} and series (\ref{f+t*}), (\ref{f-t*}) induce the 
character expansion for a number of matrix integrals. 
Relations (\ref{l1'})-(\ref{l3'}) may be used in the study of the so-called $\beta=2$ ensembles. Then (\ref{l4}) yields a link
of these ensembles with the BKP tau function \cite{KvdLbispec}.

Let
\be
V(z,T)=\sum_{m>1} \frac 1m z^m T_m
\ee
and variables $T=(T_1,T_2,\dots )$ and $\bpow = (p_1,p_2,\dots)$ are linearly dependent and related as follows
\be\label{V(z)-V(x)}
V(z,T) -c(T)  =V(x,\bpow)+V(x^{-1},\bpow)\,
\qquad c(T)=\sum_{n=1}^\infty \frac{(2n)!}{n!n!} \frac{T_{2n}}{2n}
\ee
For instance, $T_1=t_1-\sum_{n=1}^\infty (2n+1)t_{2n+1}$.

Then we have
\bp\label{Proposition-beta=2}{\em
\be\label{I-n-T}
I_n(T)=\int  \Delta(z_1,\dots,z_n)^2
\prod_{i=1}^n z_i^N e^{V(z_i,T)} d\mu(z_i)=
\ee
\bea \label{s}
&=&\sum_{h_1>\cdots > h_n\ge 0 } \,
s_{\{h \}}(T) \, \pi_{\{h\}}(N) \, 
\\ \label{ss}
&=&\sum_{h_1>\cdots > h_n\ge 0 \atop {\tilde h}_1>\cdots > {\tilde h}_n\ge 0} 
s_{\{h \}}(\bpow^{(1)})\, \pi^{00}_{\{h,{\tilde h}\}}(N)\, s_{\{{\tilde h} \}}(\bpow^{(2)})
\\
&=&\sum_{h_1>\cdots > h_n\ge 0 \atop {\tilde h}_1>\cdots > {\tilde h}_n\ge 0} \,e^{c(\bpow)}\,
o_{\{h \}}(\bpow) \, \pi^{--}_{\{h,{\tilde h}\}}(N) \, o_{\{{\tilde h} \}}({\bpow})\,
e^{c({\bpow})}
\\
&=&\sum_{h_1>\cdots > h_n\ge 0 \atop {\tilde h}_1>\cdots > {\tilde h}_n\ge 0} \,e^{c(\bpow)}\,
sp_{\{h \}}(\bpow)\, \pi^{++}_{\{h,{\tilde h}\}}(N)\, sp_{\{{\tilde h} \}}({\bpow})\,
e^{c({\bpow})}
\\
&=&\sum_{h_1>\cdots > h_n\ge 0 \atop {\tilde h}_1>\cdots > {\tilde h}_n\ge 0} \,e^{c(\bpow)}\,
o_{\{h \}}(\bpow) \, \pi^{-+}_{\{h,{\tilde h}\}} \,(N) sp_{\{{\tilde h} \}}({\bpow})\,
e^{c({\bpow})}
\eea
where $T=\bpow^{(1)}-\bpow^{(2)}$ (that is $T_m=p^{(1)}_m-p^{(2)}_m$), and
\be\label{Plucker}
\pi_{\{h\}}(N) = n!\det \left[ \pi^{00}_{n-i, h_j}(N)  \right]_{i,j=1,\dots,n}
\ee
\be
\pi^{ab}_{\{h,{\tilde h}\}}(N) =n!\det \left[ \pi^{ab}_{h_i,{\tilde h}_j}  \right]_{i,j=1,\dots,n}\,,\quad a,b=\pm,0
\ee
and the related moment matrices are
\bea  
\label{moment00'}
\pi^{00}_{i,j}(N) &=& \int z^{i+j+N} d\mu (z)
\\  \label{moment++'}
\pi^{++}_{i,j}(N)&=& \int x^{i-j}(x+x^{-1})^N
 d\mu (z(x))
\\  \label{moment--'}
\pi^{--}_{i,j}(N)&=& \int x^{i-j}(x+x^{-1})^N \frac{d\mu (z(x))}{(1-x^2)^2}
\\  \label{moment-+'}
\pi^{-+}_{i,j}(N)&=& \int x^{i-j}(x+x^{-1})^N
 \frac{d\mu (z(x))}{1-x^2}
\eea
}
\ep
Proof. We use Lemma \ref{Vand(z)} in the Appendix \ref{Rewriting Vandermond} and relations 
(\ref{V(z)-V(x)}), (\ref{f+t*})-(\ref{f-t*}). The Schur functions which are 
involved
in (\ref{f+t*})-(\ref{f-t*}) we present as ratios of the determinants (\ref{Schur-x}), at last we use the simple identity
(sometimes called Andreif identity)
\[
 \int \det \left[ \tau_i(x_j)\right]_{i,j=1,\dots,n}\det \left[ \tau_i(x_j)\right]_{i,j=1,\dots,n}\prod_{i=1}^n d\mu(x_i)=
 n!\det \left[ \int \tau_i(x) \tau_j(x)  d\mu(x)\right]_{i,j=1,\dots,n}
\]
to get moment matrices (\ref{moment00'})-(\ref{moment-+'}).

\br
Series (\ref{s}) is a special case of (\ref{ss}) where $\bpow^{(2)}=0$.
\er

 \paragraph{$\beta=2$ ensembles as the BKP tau function.} Using (\ref{l4}) in Appendix one may verify that
 \be
 I_n(T(\bpow))= n! e^{c(\bpow)}\l n| e^{\sum_{m>0} \frac 1m  p_m \alpha_m} e^{\int \frac{\psi(x^{-1})
 \psi(x)}{x^{-1}-x} d\mu(z(x)) + \sqrt{2} \int \psi(x)d\mu(x) \phi} |0\r
 \ee
Then we can apple the Wick's rule to get the Pfaffian representation
\be
 I_n(T(\bpow))=n! \Pf \left[ A_{ij}(\bpow) \right]
\ee
where, for $n$ even

\br

The integrals (\ref{I-n-T}) may be related both to  the BKP hierarchy where $\bpow$ are BKP higher times 
the one-dimensional Toda chain with higher times ${\tilde\bpow}$ , see (\ref{p-tilde{p}})

\er

\section{The character expansion of two-matrix models \label{2-MM}}

Let 
\be\label{z-tilde-z-x-tilde-x}
z=x+x^{-1}\,,\quad {\tilde z}={\tilde x}+{\tilde x}^{-1}
\ee
In case we can express the variables $\bpow$ as a linear combination of the variables $\bpow^{(1)}$ and
the variables ${\tilde\bpow}$ as a linear combination of the variables $\bpow^{(2)}$:
\[
p_m=\sum_{n=1}^\infty D_{mn}p^{(1)}_n\,,\quad  {\tilde p}_m=\sum_{n=1}^\infty {\tilde D}_{mn}p^{(2)}_n
\]
in such a way that
\be
V\left(\frac{az+b}{cz+d},\bpow^{(1)}\right)=V(x,\bpow)\,,\quad 
V\left(\frac{{\tilde a}{\tilde z}+{\tilde b}}{{\tilde c}{\tilde z}+{\tilde d}},\bpow^{(1)}\right)=V({\tilde x},{\tilde\bpow})
\ee
we can apply the same method based on (\ref{l1'})-(\ref{l2'}) in Appendix \ref{Rewriting Vandermond} to get character expansion of
\be\label{2-MM-eigenvalues}
I_n(\bpow^{(1)},\bpow^{(2)})=\int  \Delta(z_1,\dots,z_n)\Delta({\tilde z}_1,\dots,{\tilde z}_n)
\prod_{i=1}^n z_i^N e^{V({ z}_i,\bpow^{(1)})-V({\tilde z}_i,\bpow^{(2)})} d\mu(z_i,{\tilde z}_i)
\ee

For the sake of simplicity let us take  
$d\mu(z,{\tilde z})=0$ if either $z\ge 1$ or ${\tilde z}\ge 1$, in (\ref{2-MM-eigenvalues}) and in addition
\[
 V(z^{-1},\bpow^{(1)})=V(x,\bpow)\,,\quad
 V({\bar z}^{-1},\bpow^{(1)})=V({\tilde x},{\tilde\bpow})
\]
such that $p_1=p^{(1)}_1$, $p_2=p^{(1)}_2$, $p_3=p^{(1)}_3-t{(1)}_1$, $p_4=p^{(1)}_4-2t{(1)}_2$ and so on.  

\be\label{2MM}
I_n(\bpow^{(1)},\bpow^{(2)})=
\ee
\bea
&=&\sum_{h_1>\cdots > h_n\ge 0 \atop {\tilde h}_1>\cdots > {\tilde h}_n\ge 0} 
s_{\{h \}}(\bpow^{(1)})\, G^{00}_{\{h,{\tilde h}\}}\, s_{\{{\tilde h} \}}(\bpow^{(2)})
\\
&=&\sum_{h_1>\cdots > h_n\ge 0 \atop {\tilde h}_1>\cdots > {\tilde h}_n\ge 0} \,e^{c(\bpow)}\,
o_{\{h \}}(\bpow) \, G^{--}_{\{h,{\tilde h}\}}(\bpow,{\tilde\bpow}) \, o_{\{{\tilde h} \}}({\tilde\bpow})\,
e^{c({\tilde\bpow})}
\\
&=&\sum_{h_1>\cdots > h_n\ge 0 \atop {\tilde h}_1>\cdots > {\tilde h}_n\ge 0} \,e^{c(\bpow)}\,
sp_{\{h \}}(\bpow)\, G^{++}_{\{h,{\tilde h}\}}(\bpow,{\tilde\bpow})\, sp_{\{{\tilde h} \}}({\tilde\bpow})\,
e^{c({\tilde\bpow})}
\\
&=&\sum_{h_1>\cdots > h_n\ge 0 \atop {\tilde h}_1>\cdots > {\tilde h}_n\ge 0} \,e^{c(\bpow)}\,
o_{\{h \}}(\bpow) \, G^{-+}_{\{h,{\tilde h}\}}(\bpow,{\tilde\bpow}) \, sp_{\{{\tilde h} \}}({\tilde\bpow})\,
e^{c({\tilde\bpow})}
\\
&=&\sum_{h_1>\cdots > h_n\ge 0 \atop {\tilde h}_1>\cdots > {\tilde h}_n\ge 0} \,
s_{\{h \}}(\bpow^{(1)}) \, G^{0+}_{\{h,{\tilde h}\}}({\tilde\bpow}) \, sp_{\{{\tilde h} \}}({\tilde\bpow})\,
e^{c({\tilde\bpow})}
\\
&=&\sum_{h_1>\cdots > h_n\ge 0 \atop {\tilde h}_1>\cdots > {\tilde h}_n\ge 0} \,
s_{\{h \}}(\bpow^{(1)}) \, G^{0-}_{\{h,{\tilde h}\}}({\tilde\bpow}) \, o_{\{{\tilde h} \}}({\tilde\bpow})\,
e^{c({\tilde\bpow})}
\eea
where
\be
G^{ab}_{\{h,{\tilde h}\}} =\det \left[ G^{ab}_{h_i,{\tilde h}_j}  \right]_{i,j=1,\dots,n}\,,\quad a,b=\pm,0
\ee
and the related moment matrices are
\bea  \label{moment00}
G^{00}_{i,j} &=& \int z^{i+N}{\tilde z}^{j+N} d\mu (z,{\tilde z})
\\  \label{moment++}
G^{++}_{i,j}&=& 
\int  x^{i-\frac{n(n-1)}{2}}{\tilde x}^{j-\frac{n(n-1)}{2}}(x+x^{-1})^N
({\tilde x}+{\tilde x}^{-1})^N  d\mu (z(x),{\tilde z}({\tilde x}))
\\  \label{moment--}
G^{--}_{i,j}&=& 
\int  x^{i-\frac{n(n-1)}{2}}{\tilde x}^{j-\frac{n(n-1)}{2}}(x+x^{-1})^N
({\tilde x}+{\tilde x}^{-1})^N \frac{d\mu (z(x),{\tilde z}({\tilde x}))}{(1-x^2)(1-{\tilde x}^2)}
\\  \label{moment-+}
G^{-+}_{i,j}&=&
\int x^{i-\frac{n(n-1)}{2}}{\tilde x}^{j-\frac{n(n-1)}{2}}(x+x^{-1})^N
({\tilde x}+{\tilde x}^{-1})^N \frac{d\mu (z(x),{\tilde z}({\tilde x}))}{1-x^2}
\\  \label{moment0+}
G^{0+}_{i,j}&=& \int z^{i+N}{\tilde x}^{j-\frac{n(n-1)}{2}}
({\tilde x}+{\tilde x}^{-1})^N  d\mu (z,{\tilde z}({\tilde x}))
\\  \label{moment0-}
G^{0-}_{i,j}&=& \int  z^{i+N}{\tilde x}^{j-\frac{n(n-1)}{2}}
({\tilde x}+{\tilde x}^{-1})^N \frac{d\mu (z,{\tilde z}({\tilde x}))}{1-{\tilde x}^2}
\eea

{\bf Example}. 
Take ${\tilde z}={\bar z}$, $x={\bar x}$, $t^{(1)}={\bar t}^{(2)}$, $t_m={\bar t}_m$, and $N=0$.
Take also $d\mu(z,{\tilde z})=f(|z|)\frac{dzd{\tilde z}}{|1-x^2|^2}$, where $z$ and $x$ are related by 
(\ref{z-tilde-z-x-tilde-x}). Then
\be
I_n(\bpow^{(1)},\bpow^{(2)})=\sum_{\lambda} r_\lambda sp_\lambda(\bpow)sp({\bar\bpow})
\ee

\section*{Acknowledgements}
   
A.O. was supported by RFBR grant 14-01-00860. 
This work has been funded by the  Russian Academic Excellence Project '5-100'.

\appendix

\section{Appendices}

\subsection{Rewriting Vandermonde determinants \label{Rewriting Vandermond}}

We have usefull elementary

\bl\label{Vand(z)} {\em
Let
\be\label{z-x}
z=x+x^{-1}\,,\quad x=\frac z2  \pm \frac 12 \sqrt{z^2 -4} 
\ee
(Joukowsky transform).

Then
\be\label{l1'}
\prod_{1\le k < j\le n}(z_k -z_j  ) = 
\prod_{k < j\le n}(1 - x_kx_j)
 \prod_{1\le k < j\le n}(x_k^{-1}-x_j^{-1})
 \ee
 \be\label{l2'}
=  \tau_+(X|0) \prod_{1\le k < j\le n}(x_k^{-1}-x_j^{-1})
\ee
\be\label{l3'}
=  \tau_-(X|0) \prod_{1\le k < j\le n}(x_k^{-1}-x_j^{-1})
\prod_{j=1}^n \frac{1}{1-x_j^2}
\ee
And as a result
\be\label{l1}
\prod_{1\le k < j\le n}(z_k -z_j  )^2 = 
\prod_{1\le k < j\le n}(1 - x_kx_j)(1 - x_k^{-1}x_j^{-1})
 \prod_{1\le k < j\le n}(x_k-x_j)( x_k^{-1}-x_j^{-1} )
 \ee
 \be\label{l2}
=  \tau_+(X|0) \tau_+(X^{-1}|0)\prod_{1\le k < j\le n}(x_k-x_j)( x_k^{-1}-x_j^{-1} )
\ee
\be\label{l3}
=  \tau_-(X|0) \tau_-(X^{-1}|0)\prod_{1\le k < j\le n}(x_k-x_j)( x_k^{-1}-x_j^{-1} )
\prod_{j=1}^n \frac{1}{(1-x_j^2)(1-x_j^{-2})}
\ee
\be\label{l3-+}
=  \tau_-(X|0) \tau_+(X^{-1}|0)\prod_{1\le k < j\le n}(x_k-x_j)( x_k^{-1}-x_j^{-1} )
\prod_{j=1}^n \frac{1}{1-x_j^2}
\ee
\be\label{l4}
=\Delta_{2n}(x_1^{-1},x_1, \dots , x_n^{-1},x_n)\prod_{j=1}^n \frac{1}{x_j^{-1}-x_j}
\ee

}
\el

Indeed, we have
\[
 (1-x_j^{-1}x_k^{-1})(x_j-x_k)=(1-x_jx_k)(x_j^{-1}-x_k^{-1})=z_j-z_k
\]
and obtain (\ref{l1})-(\ref{l2}).

\subsection{Vertex operators \label{Vertex operators}}

Vertex operators we need are as follows
\be\label{X}
{\hat X}(L,\bpow,\lambda):=e^{\sum_{n=1}^\infty \lambda^n t_n}\lambda^{L}
e^{-\sum_{n=1}^\infty \frac{1}{n\lambda^{n}}\frac{\partial}{\partial t_n}}
\,,\quad
{\hat X}^\dag(L,\bpow,\lambda):=e^{-\sum_{n=1}^\infty \lambda^n t_n}\lambda^{-L}
e^{\sum_{n=1}^\infty \frac{1}{n\lambda^{n}}\frac{\partial}{\partial t_n}}
 \ee
\be\label{Y}
{\hat Y}(L,{\bf s},\lambda):=e^{-\sum_{n=1}^\infty \lambda^{-n} s_n}\lambda^{L}
e^{\sum_{n=1}^\infty \frac{\lambda^{n}}{n}\frac{\partial}{\partial s_n}}
\,,\quad
{\hat Y}^\dag(L,{\bf s},\lambda):=e^{\sum_{n=1}^\infty \lambda^{-n} s_n}\lambda^{-L}
e^{-\sum_{n=1}^\infty \frac{\lambda^{n}}{n}\frac{\partial}{\partial s_n}}
\ee
Interesting historical fact that the formula which relates fermions to bosons first was found in \cite{PogrebkovSushko}.

The following bosonization relation is useful
 \be\label{p-bosonization}
 \l L+N|\Gamma(\bpow+\sum_{i=1}^N[p_i])=
 \frac{\l L| \psi^\dag(p_1^{-1})\cdots  \psi^\dag(p_N^{-1})\g_+(\bpow)}{\prod_{i=1}^N p_i^{(L+1)(N-1)} \prod_{i>j}(p_i-p_j)}
 \ee

  Introduce
   \be
   {\hat \Omega}_n(L,\bpow)=\res_\lambda \left(\frac{\partial^n {\hat X}(L,\bpow,\lambda)}{\partial\lambda^n} 
   {\hat X}^\dag(L,\bpow,\lambda)\right)\,,\quad
   {\hat \Omega}_n^*(L,{\bf s})=\res_\lambda  \left({\hat Y}^\dag(L,{\bf s},\lambda) \frac{\partial^n 
   {\hat Y}(L,{\bf s},\lambda)}{\partial\lambda^n}\right)
   \ee

\subsection{Fermions \label{Fermions}}

We shall remind some facts and notations of \cite{JM}. Introduce free fermionic fields
$\psi(z)=\sum_{i\in\mathbb{Z}} \psi_i z^i$, $\psi^\dag(z)=\sum_{i\in\mathbb{Z}} \psi^\dag_{-i-1} z^i$ whose Fourier components 
anti-commute as follows
$\psi_i\psi_j+\psi_j\psi_i=\psi^\dag_i\psi^\dag_j+\psi^\dag_j\psi^\dag_i=0$ and 
$\psi_i\psi^\dag_j+\psi^\dag_j\psi_i=\delta_{i,j}$ where $\delta_{i,j}$ is the Kronecker symbol. We put
 \be
\psi_i|0\r=\psi^\dag_{-i}|0\r =\l 0|\psi_{-i}=\l 0|\psi^\dag_{i}=0
 \ee
where $\l 0|$ and $|0\r$ are left and right vacuum vectors of the fermionic Fock space, $\l 0|\cdot 1 \cdot |0\r=1$. 
Also introduce
 \be
\l n|=
\cases{
\l 0|\psi^\dag_{0}\cdots \psi^\dag_{n-1}\quad {\rm if }\quad n > 0 \cr
\l 0|\psi_{-1}\cdots \psi_{-n}\quad {\rm if }\quad n < 0 
 }
\,,\quad
|n \r =
\cases{
\psi_{n-1}\cdots \psi_{0}|0\r\, \quad {\rm if }\quad n > 0 \cr
\psi^\dag_{-n}\cdots \psi^\dag_{-1}|0\r\, \quad {\rm if }\quad n < 0 
 }
 \ee
Then $\l n|\cdot 1 \cdot |m\r=\delta_{n,m}$.

Following \cite{KvdLbispec} we introduce an additional Fermi mode which we shall denote by 
$\phi$ with properties
\footnote{In notations of \cite{KvdLbispec} our $\psi_n$, $\psi_n^\dag$ and $\phi$ read respectively as
$\psi_{n+\frac 12}$, $\psi^\dag_{n+\frac 12}$ and $\psi_0$} 
 \be
\phi\psi_i+\psi_i\phi=\phi\psi_i^\dag+\psi_i^\dag\phi=0,\quad \phi^2=\frac 12
 \ee
 \be
\phi|0\r=|0\r\frac{1}{\sqrt 2} ,\qquad \l 0|\phi=\frac{1}{\sqrt 2}\l 0|
 \ee
such that $\l L|\phi |L\r=\frac{(-)^L}{\sqrt{2}}$.
 
Now, two-component Fermi fields used in Section \ref{2-MM} are defined as
  \be
  \psi^{(i)}(z)=\sum_{n\in\mathbb{Z}} z^n\psi_{2n+i}\,,\quad  \psi^{(i)\dag}(z)=
  \sum_{n\in\mathbb{Z}} z^{-n-1}\psi_{2n+i}^\dag
  \ee
where $i=1,2,3$.  Other details about multi-component fermions may be found in \cite{JM}, and in relation to matrix 
models in \cite{HO-2006}.

We have
 \be
{\hat  X}(L,\bpow,\lambda)X^\dag(L,\mu)\l N+L|\Gamma(\bpow)g\Gamma^\dag({\bf s})|L\r=
 \l N+L|\Gamma(\bpow)\psi(\lambda)\psi^\dag(\mu)g\Gamma^\dag({\bf s})|L\r
 \ee
 \be
{\hat Y}^\dag(L,{\bf s},\mu) Y(\lambda)\l N+L|\Gamma(\bpow)g\Gamma^\dag({\bf s})|L\r=
\l N+L|\Gamma(\bpow)g\psi(\lambda)\psi^\dag(\mu)\Gamma^\dag({\bf s})|L\r
 \ee

 Then it follows that
  \be
  {\hat \Omega}_n(L,\bpow)\l N+L|\Gamma(\bpow)g\Gamma^\dag({\bf s})|L\r=
  \l N+L|\Gamma(\bpow){\tilde \Omega}_n g\Gamma^\dag({\bf s})|L\r
  \ee
  \be
  {\hat \Omega}_n(L,\bpow)\l N+L|\Gamma(\bpow)g\Gamma^\dag({\bf s})|L\r=
\l N+L|\Gamma(\bpow)g \tilde {\Omega}_n\Gamma^\dag({\bf s})|L\r
  \ee
  where
  \be
  {\tilde \Omega}_n=\res_\lambda \left(\frac{\partial^n \psi(\lambda)}{\partial\lambda^n} \psi^\dag(\lambda)\right)
  \ee
Using the fermionic representation  one may verify that tau functions related to the considered ensembles (dd-OE, dd-GinOE,
dd-SE, dd-GinSE) obey the constraints
   \be
   \left({\hat \Omega}_n(L,\bpow)-{\hat \Omega}^*_n(L,{\bf s})\right)\tau(L,\bpow',{\bf s}')=0,\quad n\ge 1 \quad {\rm odd}
   \ee
   where $t'_k=t_k-\frac 12 \delta_{2,k}$ $s'_k=s_k-\frac 12 \delta_{2,k}$ 
   (this shift appears due to the Gauss measure in undeformed ensembles).

    \subsection{The Schur function\label{Schur-f}}
    
    Consider polynomials $h_n(\bpow)$ defined by $e^{\sum_{n=1}^\infty z^nt_n}=\sum_{n=0}^\infty z^nh_n(\bpow)$. Then the Schur
    function labeled by a partition $\lambda=(\lambda_1,\dots,\lambda_k>0)$ may be defined as
     $\label{Schur}
     s_\lambda(\bpow)=\det \left( h_{\lambda_i-i+j}(\bpow)\right)_{i,j=1,\dots,k}
     $.
     The notation $s_\lambda(X)$ denotes (\ref{Schur}) where $\bpow=\sum_i [x_i]$ where $x_i$ are eigenvalues of $X$.

\subsection{Characters of classical Lie groups/algebras \label{Characters of}}
We recall some information about characters of classical Lie algebras as it is presented in  \cite{FH} Chapter 24.
The character of a simple lie algebra is given by the Weyl character formula. For a dominant integral weight $\lambda$ 
the character $ch_\lambda$ is equal to
\be
ch_\lambda=\frac{A_{\lambda+\rho}}{A_{\rho}}, \quad\mbox{where } A_\mu=\sum_{w\in W} {\sgn(w)} e(w(\mu)).
\ee
Here $W$ is the Weyl group of the simple Lie algebra. and $\rho$ the sum of all the fundamental weights, or equivalently 
half the sum of all positive roots $R^+$.
One has
\[
 A_\rho=\prod_{\alpha\in R^+} (e(\alpha/2)-e(-\alpha/2))=
e(\rho)\prod_{\alpha\in R^+} (1-e(-\alpha))=\prod_{\alpha\in R^+} e(-\rho)(e(\alpha)-1)
\]
\paragraph{First case: $sl_n$ or rather $gl_n$.}
We identify the standard Cartan subalgebra, the diagonal matrices of $gl_n$, with its dual via the trace form 
$(a,b)=\mbox{trace}\, (ab)$.
Let $\epsilon_i=E_{ii}$, and assume $\lambda=\sum_{i=1}^n\lambda_i\epsilon_i $.
In this case $R^+$ consists of all elements $\epsilon_i-\epsilon j$, with $i<j$ and 
\[
\rho=\frac12 \sum_{i<j}(\epsilon_i-\epsilon_j)=
\sum_{i=1}^n (\frac{n+1}2-i)\epsilon_i
\]
N.B. Fulton and Harris \cite{FH} have a different formula for $\rho$, they claim $\rho=\sum_i(n-i)\epsilon_i$ which is wrong.
 The Weyl group is the group $S_n$, the permutation group that permutes the elements $\epsilon_i$. Denote by $x_i=e(\epsilon_i)$,
 then
\be
\label{A lambda-rho}
A_{\lambda+\rho}=\sum_{w\in S_n} {\sgn (w)} x_{w(1)}^{\lambda_1+\frac{n+1}2-1}x_{w(2)}^{\lambda_2+\frac{n+1}2-2}\cdots
x_{w(n)}^{\lambda_n\frac{n+1}2-n}= \det \left[ x_j^{\lambda_i+\frac{n+1}2-i}\right]_{1\le i,j\le n}
\ee
Taking $\lambda=0$ in (\ref{A lambda-rho}), we obtain $A_{\rho}=\det \left[ x_j^{\frac{n+1}2-i}\right]$.
Thus
\[
 ch_\lambda=s_\lambda=
\frac{\det \left[ x_j^{\lambda_i+\frac{n+1}2-i}\right]_{1\le i,j\le n}}{\det \left[ x_j^{\frac{n+1}2-i}\right]_{1\le i,j\le n}}
=
\frac{\det \left[ x_j^{\lambda_i+n-i}\right]_{1\le i,j\le n}}{\det \left[ x_j^{n-i}\right]_{1\le i,j\le n}}
\]
Another presentation of the character is the so called Giambelli or determinantal formula. One expresses the character in the elementary Schur functions functions 
$S_{(k)}(x)$  defined in (\ref{Schur}):
\[
 s_\lambda(x)=\det \left[s_{(\lambda_i+j-i)}(x)\right]_{1\le i,j\le n}
\]

\paragraph{Second case: $sp_{2n}$.}
The positive roots aren now $\epsilon_i-\epsilon_j$, with $1\le i<j\le n$ and $\epsilon_i+\epsilon_j$ with $i\le j$. The element
$\rho=\sum_{i=1}^n (n+1-i)\epsilon_i$ and the Weyl group is the group that permutes all $\epsilon i$ allowing also all possible 
sign changes, i.e., $\epsilon_i\mapsto \pm \epsilon_j$., hence it is the semi-direct product of $S_n$ and $\mathbb{Z}_2^n$.
Then
\[
 A_\mu=\sum_{w\in S_n} \sum_{\sigma\in\mathbb{Z}_2^n} {\sgn(w)\sgn(\sigma)} e(\sum_i (-)^{\sigma_i} \mu_i w(\epsilon_i)).
\]
Here $\sigma=(\sigma_1,\ldots\sigma_n)$ and $\sgn(\sigma)=(-)^{\sum_i \sigma_i}$. Using again $x_i=e(\epsilon_i)$, it is 
straightforward to check 
\be
\label{A mu}
 A_\mu=\sum_{w\in S_n} \sgn(w)(x^{\mu_i}_{\sigma(i)}-x^{-\mu_i}_{\sigma(i)})
\ee
Hence
\[
 ch_\lambda=sp_\lambda=
 \frac{\det \left[ x_j^{\lambda_i+n-i+1}-x_j^{-\lambda_i-n+i-1}\right]_{1\le i,j\le n}}
 {\det \left[ x_j^{n-i+1}-x_j^{-n+i-1}\right]_{1\le i,j\le n}}
\]
The  determinantal formula in this case is due to Koike en Terada \cite{KoikeTerada}
\[
sp_\lambda (x)= \frac12\det \left[ s_{(\lambda_i-i+j)}(x)+s_{(\lambda_i-i-j+2)}(x)\right]_{i,j=1,\dots,n}
\]
where we substitute $x_{n+i}=x_i^{-1}$
\paragraph{Third case: $so_{2n+1}$.}
The positive roots aren now $\epsilon_i-\epsilon_j$, with $1\le i<j\le n$ and $\epsilon_i+\epsilon_j$ with $i< j$ and all 
$\epsilon_i$. The element $\rho=\sum_{i=1}^n (n+\frac12-i)\epsilon_i$. The Weyl group is the same as in the $sp_{2n}$ case. 
Hence, also $A_\mu$ is the same, viz. (\ref{A mu}), as for $sp_{2n}$.
Hence 
\[
 ch_\lambda=o_\lambda=\frac{\det \left[ x_j^{\lambda_i+n-i+\frac12}-x_j^{-\lambda_i-n+i-\frac12}\right]_{1\le i,j\le n}}
 {\det \left[ x_j^{n-i+\frac12}-x_j^{-n+i-\frac12}\right]_{1\le i,j\le n}}
=
\frac{\det \left[ x_j^{\lambda_i+n-i+1}-x_j^{-\lambda_i-n+i}\right]_{1\le i,j\le n}}
{\det \left[ x_j^{n-i+1}-x_j^{-n+i}\right]_{1\le i,j\le n}}
\]
The determinantal formula is equal to 
\be
\label{olambda}
o_\lambda (x)= \det \left[ s_{(\lambda_i-i+j)}(x)-s_{(\lambda_i-i-j)}(x)\right]_{i,j=1,\dots,n}
\ee
here $x_{n+i}=x_i^{-1}$ and $x_{2n+1}=1$. In term of irreducible charcters $s_{(k)}^\circ (x)=s_{(k)}(x)-s_{(k-2)}(x)$, one can rewrite (\ref{olambda}) to
\be
\label{olambda2}
o_\lambda (x)= \frac12\det \left[ s_{(\lambda_i-i+j)}^\circ (x)+s_{(\lambda_i-i-j+2)}^\circ(x)\right]_{i,j=1,\dots,n}
\ee

\paragraph{Fourth case: $so_{2n}$.}
The positive roots aren now $\epsilon_i-\epsilon_j$, with $1\le i<j\le n$ and $\epsilon_i+\epsilon_j$ with $i< j$ and $\rho$ 
is the same as for $gl_n$, viz. $\rho=\sum_{i=1}^n (n-i)\epsilon_i$. However the Weyl group is a subgroup of the Weyl group of 
$so_{2n+1}$, one only allows an even number of sign changes. This leads to
\[
 A_\mu=\frac12 \left(\det\left[ x_j^{\mu_i}-x_j^{-\mu_i}\right]_{1\le i,j\le n}+
 \det\left[ x_j^{\mu_i}+x_j^{-\mu_i}\right]_{1\le i,j\le n}\right)
\]
and hence
\[
 ch_\lambda=\frac{\det\left[ x_j^{\lambda_i+n-i}-x_j^{-\lambda_i-n+i}\right]_{1\le i,j\le n}+
 \det\left[ x_j^{\lambda_i+n-i}+x_j^{-\lambda_i-n+i}\right]_{1\le i,j\le n}}
 {\det\left[ x_j^{n-i}+x_j^{-n+i}\right]_{1\le i,j\le n}}
\]
The determinantal formulas (\ref{olambda}) and (\ref{olambda2}) also hold in this case with the restriction that in this case 
$x_{n+i}=x_i^{-1}$, the  element $x_{2n+1}$ does not exist.

 \subsection{Pfaffians\label{Pfaffians}}

If $A$ an anti-symmetric matrix of an odd order its determinant
vanishes. For even order, say $k$, the following multilinear form
in $A_{ij},i<j\le k$
 \be\label{Pf''}
\Pf [A] :=\sum_\sigma
{\sgn(\sigma)}\,A_{\sigma(1),\sigma(2)}A_{\sigma(3),\sigma(4)}\cdots
A_{\sigma(k-1),\sigma(k)}
 \ee
where sum runs over all permutation restricted by
 \be
\sigma:\,\sigma(2i-1)<\sigma(2i),\quad\sigma(1)<\sigma(3)<\cdots<\sigma(k-1),
 \ee
 coincides with the square root of $\det A$ and is called the
 {\em Pfaffian} of $A$, see, for instance \cite{Mehta}. As one can see the Pfaffian  contains
 $1\cdot  3\cdot 5\cdot \cdots \cdot(k-1)=:(k-1)!!$ terms.
 
 \paragraph{ Wick's relations.} Let each of $w_i$ be a linear
combination of Fermi operators:
  \[
{\hat w}_i=\sum_{m\in\mathbb{Z}}\,v_{im}\psi_m\,+\,
\sum_{m\in\mathbb{Z}}\,u_{im}\psi^\dag_m\,
 ,\quad i=1,\dots,n
 \]
  Then the Wick's formula is
  \be\label{Wick} \l l|{\hat w}_1\cdots {\hat w}_n |l\r =
\cases{
\Pf\left[ A \right]_{i,j=1,\dots,n} \quad {\rm if\,\,n\,is\,even}  \cr
0 \qquad\qquad\qquad\quad\, \mbox{otherwise}
 }
  \ee  
    where $A$ is $n$ by $n$ antisymmetric matrix with entries
 $ A_{ij}\, = \,\l l|{\hat w}_i {\hat w}_j|l\r\, ,\quad i<j$.

 \subsection{Hirota equations \label{Hirota}}
 
 KP Hirota equation:
 \be
  \oint\frac{dz}{2\pi i} e^{V(\bpow'-\bpow,z)}
  \tau_\mp(\bpow'-[z^{-1}]|\bpow^*)
  \tau_\mp(\bpow+[z^{-1}]|\bpow^*) =0
 \ee
 For $\tau_\mp$ defined by (\ref{f1})  we have for the left hand side
 \[
  \oint\frac{dz}{2\pi i} \left(1-z^{-2} \right) e^{\sum_{k=1}^\infty \frac 1k p_k'
  \left( z^k +z^{-k} \right)}
  e^{-\sum_{k=1}^\infty \frac 1k p_k
  \left( z^k +z^{-k} \right)}  
 \]
 which is equal to zero. Thus, $\tau_\mp(\bpow | \bpow^*)$ is the KP tau function with respect to the $\bpow$ variables.

 Now in view of Giambelli relations for characters \cite{Baker}
 \be
 o_{\lambda}(\bpow)=\det \left( o_{(\alpha_i|\beta_j)}(\bpow)\right)\,,\qquad 
 sp_{\lambda}(\bpow)=\det \left( sp_{(\alpha_i|\beta_j)}(\bpow)\right)\,,\qquad \lambda=(\alpha|\beta)
 \ee
  we can write
 \be\label{f-tau-KP-t}
 \tau_-(\bpow |\bpow^*)=\l 0|e^{\sum_{k>0} \frac 1k J_kp_k} e^{\sum_{i,j\ge 0} (-)^j o_{(i|j)}(\bpow^*)\psi_i\psi^\dag_{-1-j}} |0\r
 \ee
 \be\label{f+tau-KP-t}
 \tau_+(\bpow |\bpow^*)=\l 0|e^{\sum_{k>0} \frac 1k J_kp_k} e^{\sum_{i,j\ge 0} (-)^j sp_{(i|j)}(\bpow^*)\psi_i\psi^\dag_{-1-j}} |0\r
 \ee
  
 Hirota equations for the large BKP hierarchy were written in \cite{KvdLbispec}. For 2-BKP hierarchy 
 Hirota equations are as follows \cite{OST-I}
 \bea\label{Hirota-2lBKPtau}
  \oint\frac{dz}{2\pi i}z^{N'+l'-N-l-2}e^{V(\bpow'-\bpow,z)}
  \tau_{N'-1}(l',\bpow'-[z^{-1}],{\bf s}')
  \tau_{N+1}(l,\bpow+[z^{-1}],{\bf s}) \nonumber\\
+ \oint\frac{dz}{2\pi i}z^{N+l-N'-l'-2}e^{V(\bpow-\bpow',z)}
  \tau_{N'+1}(l',\bpow'+[z^{-1}],{\bf s}')
  \tau_{N-1}(l,\bpow-[z^{-1}],{\bf s}) \nonumber\\
= \oint\frac{dz}{2\pi i}z^{l'-l}e^{V({\bf s}'-{\bf s},z^{-1})} 
  \tau_{N'-1}(l'+1,\bpow',{\bf s}'-[z])
  \tau_{N+1}(l-1,\bpow,{\bf s}-[z]) \nonumber \\
+ \int\frac{dz}{2\pi i}z^{l-l'}e^{V({\bf s}'-{\bf s},z^{-1})}
  \tau_{N'+1}(l'-1,\bpow',{\bf s}'+[z])
  \tau_{N-1}(l+1,\bpow,{\bf s}+[z]) \nonumber\\
+ \frac{(-1)^{l'+l}}{2}(1-(-1)^{N'+N})
  \tau_{N'}(l',\bpow',{\bf s}')\tau_N(l,\bpow,{\bf s}) 
\eea
 The difference Hirota equation may be obtained from the previous one \cite{OST-I}
 \bea
  - \frac{\beta}{\alpha-\beta}
    \tau_N(l,\bpow+[\beta^{-1}])\tau_{N+1}(l,\bpow+[\alpha^{-1}]) 
  - \frac{\alpha}{\beta-\alpha}
    \tau_N(l,\bpow+[\alpha^{-1}])\tau_{N+1}(l,\bpow+[\beta^{-1}]) 
  \nonumber\\
  + \frac{1}{\alpha\beta}
    \tau_{N+2}(l,\bpow+[\alpha^{-1}]+[\beta^{-1}])\tau_{N-1}(l,\bpow) 
  = \tau_{N+1}(l,\bpow+[\alpha^{-1}]+[\beta^{-1}])\tau_N(l,\bpow). 
\eea

\subsection{Integrals over the unitary, orthogonal and symplectic groups \label{Haar-integration}} 

For $\mathbb{U}(n)$ the Haar measure is
\be\label{Haar-unitary}
 d_*U =\frac{1}{(2\pi )^n}  
\prod_{1\le j<k\le n}\vert e^{i\theta_j}-e^{-i\theta_k} \vert ^2 
 \prod_{j=1}^n d\theta_j\,,\quad -\pi \le \theta_1<\dots\theta_n\le \pi
 \ee
 where $e^{\theta_1},\dots,e^{\theta_n}$ are eigenvalues of $U\in\mathbb{U}(n)$.

For $\mathbb{O}(2n)$ the Haar measure is
 \be\label{Haar-even-othogonal'}
 d_*O=\frac{2^{(n-1)^2}}{\pi^n } 
\,\prod_{i<j}^n\,(\cos \theta_i -\cos \theta_j )^2 \,\prod_{i=1}^n d\theta_i\,,
\quad 0 \le \theta_1 \le \cdots \le \theta_n \le \pi
 \ee
where $e^{\theta_1},e^{-\theta_1},\dots,e^{\theta_n},e^{-\theta_n}$ are eigenvalues of $O\in\mathbb{O}(2n)$.

For $\mathbb{O}(2n+1)$ the Haar measure is
 \be\label{Haar-odd-othogonal'}
d_*O=\frac{2^{n^2}}{\pi^n } 
\,\prod_{i<j}^n\,(\cos \theta_i -\cos \theta_j )^2 \,
\prod_{i=1}^n \sin^2\frac{\theta_i}{2}d\theta_i\,,\quad
0\le \theta_1 \le \cdots \le \theta_n \le \pi
 \ee
where $e^{\theta_1},e^{-\theta_1},\dots,e^{\theta_n},e^{-\theta_n},1$ are eigenvalues of $O\in\mathbb{U}(2n+1)$.

For $\mathbb{S}p(2n)$ the Haar measure is
 \be\label{Haar-symplectic'}
 d_*S=\frac{2^{n^2}}{\pi^n } 
\,\prod_{i<j}^n\,(\cos \theta_i -\cos \theta_j )^2 \,
\prod_{i=1}^n \sin^2 {\theta_i}d\theta_i\,,\quad
0\le\theta_1 \le \cdots \le \theta_n \le \pi
 \ee
where $e^{\theta_1},e^{-\theta_1},\dots,e^{\theta_n},e^{-\theta_n}$ are eigenvalues of $S\in\mathbb{S}p(2n)$.

 \subsection{Characters as vacuum expectation values  \label{characters}}
We can write the determinatal expression (\ref{Schur-t}) for $s_\lambda(\bpow)$ as a (vacuum) expectation value \cite{JM}
\be
\label{Schur-t2}
s_\lambda(\bpow)=\langle k|\Gamma(\bpow)\psi_{k+\lambda_1-1}\psi_{k+\lambda_2-2}\cdots \psi_{k+\lambda_n-n}|k-n\rangle
\ee
Now using the commutation relations of $\Gamma(\bpow)$ with the fermions $\psi_j$, we define
\be
\label{Schur-t3} 
\psi_j(\bpow):=\Gamma(\bpow)\psi_j\Gamma(\bpow)^{-1}=\sum_{i=0}^\infty s_{(i)}(\bpow)\psi_{j-i},\qquad
\psi_j^\dagger(\bpow):=\Gamma(\bpow)\psi_j^\dagger\Gamma(\bpow)^{-1}=\sum_{i=0}^\infty s_{(i)}(-\bpow)\psi_{j+i}^\dagger
\ee
Hence,
\be
\label{Schur-t4}
s_\lambda(\bpow)=\langle k|\psi_{k+\lambda_1-1}(\bpow)\psi_{k+\lambda_2-2}(\bpow)\cdots 
\psi_{k+\lambda_n-n}(\bpow)|k-n\rangle
\ee
which is exactly the determinant of (\ref{Schur-t}).
On the other hand, if we write $\lambda$ in the Frobenius notation $\lambda=(a_1, a_2,\ldots ,a_r|b_1,b_2,\ldots ,b_r)$ where 
$a_i=\lambda_i-i\ge 0$ for $1\le i\le r$ and $\lambda_{r+1}-r-1<0$, and $b_i=\lambda_i'-i\ge 0$ for $1\le i\le r$ and 
$\lambda_{r+1}'-r-1<0$ for $\lambda'$ the conjugate partition of $\lambda$. If we drop the condition that 
$a_i=\lambda_i-i\ge 0$ and $b_i=\lambda_i'-i\ge 0$, then we can see 
\[
 \lambda=(a_1, a_2,\ldots |b_1,b_2,\ldots )\qquad\mbox{with }a_i=\lambda_i-i,\ \mbox{and } b_i=\lambda_i'-i\
\]
as two infinite sequences. It is then well known (see e.g. \cite{Samra-King}) that the sets
\[ 
A=\{ a_1,a_2,a_3,\ldots\}\quad\mbox{and } B=\{ -b_1-1,-b_2-1,-b_3-1,\ldots\}
\]
form a partition of $\mathbb{Z}$, thus
\be
\label{Schur-t5}
\psi_{k+\lambda_1-1}\psi_{k+\lambda_2-2}\cdots \psi_{k+\lambda_n-n}|k-n\rangle=(-)^{b_1+\cdots +b_r}
\psi_{k+a_1}\psi_{k+a_2}\cdots \psi_{k+a_r}
\psi_{k-b_r-1}^\dagger\psi_{k-b_{r-1}-1}^\dagger\cdots \psi_{k-b_1-1}^\dagger
|k\rangle
\ee
and 
\be
\label{Schur-t6}
s_\lambda(\bpow)=(-)^{b_1+\cdots +b_r+\frac{r(r-1)}2}
\langle k|\psi_{k+a_1}(\bpow)\psi_{k+a_2}(\bpow)\cdots \psi_{k+a_r}(\bpow)
\psi_{k-b_1-1}^\dagger(\bpow)\psi_{k-b_2-1}^\dagger(\bpow)\cdots \psi_{k-b_r-1}^\dagger(\bpow)|k\rangle
\ee
Now using Wick's relation, see Appendix,
this is equal to the Pfaffian expression
\be
\label{Schur-t7}
(-)^{b_1+\cdots +b_r+\frac{r(r-1)}2}\Pf\left[ A_\lambda \right]_{i,j=1,\dots,2r},\quad \mbox{ where }
A_\lambda= \left(
\begin{array} {cc}0 & B \\
-B^{tr} & 0 \end{array}
\right),\quad \mbox{ for } B_{ij}=\langle k|\psi_{k+a_i}(\bpow)\psi_{k-b_j-1}^\dagger(\bpow)|k\rangle
\ee
Now 
\[
 \Pf\left[ A_\lambda \right]_{i,j=1,\dots,2r}
=(-)^{\frac{r(r-1)}2} \det \left[B \right]_{i,j=1,\dots,r}
\]
and 
\[
B_{ij}=\langle k|\Gamma(\bpow)\psi_{k+a_i}\psi_{k-b_j-1}^\dagger|k\rangle= (-)^{b_j}s_{(a_i|b_j)}(\bpow)
\]
 Thus
\be
\label{Schur-t8}
s_\lambda(\bpow)=\det \left[ s_{(a_i|b_j)}(\bpow)\right]_{i,j=1,\dots,r}
\ee
which is a well known result of  Littlewood \cite{Littlewood} (see also \cite{Samra-King}).

The orthogonal and symplectic characters can also be expressed as determinants (see e.g.  \cite{KoikeTerada}, 
Theorem 1.3.2 and 1.3.3)
\be
\label{Schur-t9}
o_\lambda (\bpow)= \det \left[ s_{(\lambda_i-i+j)}(\bpow)-s_{(\lambda_i-i-j)}(\bpow)\right]_{i,j=1,\dots,n},\qquad
sp_\lambda (\bpow)= \frac12\det \left[ s_{(\lambda_i-i+j)}(\bpow)+s_{(\lambda_i-i-j+2)}(\bpow)\right]_{i,j=1,\dots,n}
\ee
and similarly one has a formula ala (\ref{Schur-t8}), see\cite{Samra-King}
\be
\label{Schur-t10}
o_\lambda(\bpow)=\det \left[ o_{(a_i|b_j)}(\bpow)\right]_{i,j=1,\dots,r}
\quad\mbox{ and } sp_\lambda(\bpow)=\det \left[ sp_{(a_i|b_j)}(\bpow)\right]_{i,j=1,\dots,r}
\ee
Using (\ref{Schur-t10}), we calculate $o_{(a+1, 1^b)}(\bpow)=o_{(a|b)}(\bpow)$ explicitly
\be
\label{Schur-t11}
o_{(a|b)}=\det \left[
\begin{array}{ccccc}s_{(a+1)}-s_{(a-1)}&s_{(a+2)}-s_{(a-2)}&s_{(a+3)}-s_{(a-3)}&\cdots&s_{(a+b+1)}-s_{(a-b-1)}\\
 s_{(0)}&s_{(1)}&s_{(2)}&\cdots&s_{(b)}\\
0&s_{(0)}&s_{(1)}&\cdots&s_{(b-1)}\\
\vdots&\vdots&\vdots&\ddots&\vdots\\
0&0&0&\cdots&s_{(1)}
\end{array}
\right]
\ee
\[
 =s_{(a|b)}-s_{(a-2|b)}+\det
\left[
\begin{array}{ccccc}0&s_{(a)}-s_{(a-2)}&s_{(a+1)}-s_{(a-3)}&\cdots&s_{(a+b-1)}-s_{(a-b-1)}\\
 s_{(0)}&s_{(1)}&s_{(2)}&\cdots&s_{(b)}\\
0&s_{(0)}&s_{(1)}&\cdots&s_{(b-1)}\\
\vdots&\vdots&\vdots&\ddots&\vdots\\
0&0&0&\cdots&s_{(1)}
\end{array}
\right]
\]
\[
 =s_{(a|b)}-s_{(a-2|b)}-o_{(a-1|b-1)}=\sum_{j=0}^{b} (-)^j s_{(a-j|b-j)}-s_{(a-j-2|b-j)}
\]

A similar calculation, using again (\ref{Schur-t10}), shows that $sp_{((a+1, 1^b)}(\bpow)=sp_{(a|b)}(\bpow)$ is equal to
\be
\label{Schur-t12}
sp_{(a|b)}=s_{(a|b)}-s_{(a-1|b-1)}-o_{(a|b-2)}=\sum_{j=0}^{b} (-)^j s_{(a-j|b-j)}-s_{(a-j|b-j-2)}
\ee
which is also equal to
\be
\label{Schur-t13}
sp_{(a|b)}=\sum_{kj0}^{a} (-)^j s_{(a-j|b-j)}-s_{(a-j|b-j-2)}
\ee
We can write both (\ref{Schur-t11}) and (\ref{Schur-t13}) as vacuum expectation values, viz.,
\be
\label{Schur-t14}
o_{(a|b)} (\bpow)=(-)^b\sum_{j=0}^\infty
\langle k|\Gamma(\bpow)(\psi_{k+a-j}- \psi_{k+a-j-2})\psi_{k-b+j-1}^\dag|k\rangle
\ee
\[
=(-)^b\mbox{Res}_{z=0}\mbox{Res}_{w=0}\, \frac{(z^{-a-1}-z^{-a+1})w^{-b-1}}{1-zw}\left( \frac{w}{z}\right)^k
\langle k|\Gamma(\bpow)\psi(z)\psi^\dag(w)|k\rangle\, ,
\]
where 
\be
\label{fermionicfields}
\psi(z)=\sum_{j\in\mathbb{Z}}\psi_jz^j,\qquad \psi^\dag(z)=\sum_{j\in\mathbb{Z}}\psi^\dag_jz^{-j-1}
\ee
And 
\be
\label{Schur-t15}
sp_{(a|b)} (\bpow)=(-)^b\sum_{j=0}^\infty
 \langle k|\Gamma(\bpow)\psi_{k+a-j}(\psi_{k-b+j-1}^\dag-\psi_{k-b+j+1}^\dag)|k\rangle
\ee
\[
=(-)^b\mbox{Res}_{z=0}\mbox{Res}_{w=0}\, \frac{z^{-a-1}(w^{-b-1}-w^{-b+1})}{1-zw}\left( \frac{w}{z}\right)^k
\langle k|\Gamma(\bpow)\psi(z)\psi^\dag(w)|k\rangle
\]
Now let $\lambda=(a_1,a_2,\cdots a_r|b_1,b_2,\cdots b_r)$, then 
\be
\label{Schur-t16}
o_{\lambda} (\bpow)=
(-)^{\sum b_j}\prod_{i=1}^r
\mbox{Res}_{z_i=0}\mbox{Res}_{w_i=0}
(z_i^{-a_i-1}-z_i^{-a_i+1})w_i^{-b_i-1}
\det\left[\left( \frac{w_j}{z_i}\right)^k
\langle k|\Gamma(\bpow)\frac{\psi(z_i)\psi^\dag(w_j)}{1-z_iw_j}|k\rangle\right]_{i,j=1,\dots,r}
\ee

And similarly
\be
\label{Schur-t17}
sp_{\lambda} (\bpow)=(-)^{\sum b_j}\prod_{i=1}^r
\mbox{Res}_{z_i=0}\mbox{Res}_{w_i=0}
z_i^{-a_i-1}(w_i^{-b_i-1}-w_i^{-b_i+1})
\det\left[\left( \frac{w_j}{z_i}\right)^k
\langle k|\Gamma(\bpow)\frac{\psi(z_i)\psi^\dagger(w_j)}{1-z_iw_j}|k\rangle\right]_{i,j=1,\dots,r}
\ee
  
\end{document}